# Per-Lesion Radiomics Analysis of [68Ga]-DOTA FAPI-46 and [18F]-FDG PET/CT in Non–Small Cell Lung Cancer: A Pilot Study

Setareh Hasanabadi MSc[1], Maryam Cheraghi MSc[1,2], Hossein Behnam Manesh PhD[2,3], Mohadeseh Bayat[4], Mehrdad Bakhshayesh Karam MD[2,5,6], Andrea Corsi MD[7], Yazdan Salimi PhD[8], Mohammad Saber Azimi MSc[1], Abtin Doroudinia MD[9], Arezu Karami MD[10], Hossein Arabi PhD[8†]

[1]Department of Medical Radiation Engineering, Shahid Beheshti University, Tehran, Iran

[2]Chronic Respiratory Diseases Research Center, National Research Institute of Tuberculosis and Lung Diseases, Shahid Beheshti University of Medical Sciences, Tehran, Iran;

[3]Department of Pharmaceutical Chemistry and Radiopharmacy, Faculty of Pharmacy, Shahid Beheshti University of Medical Sciences, Tehran, Iran

[4]Department of Nuclear Medicine, School of Medicine, Shahid Beheshti University of Medical Sciences, Taleghani Hospital, Tehran, Iran

[5]Shahid Beheshti University of Medical Sciences, Radiology Department, Tehran, Iran

[6]Masih Daneshvari Hospital, PET/CT Division, Tehran, Iran

[7]Radiomics.bio, LégiaPark, Boulevard Patience et Beaujonc 3, 4000 Liège, Belgium

[8]Division of Nuclear Medicine & Molecular Imaging, Geneva University Hospital, CH-1211,

Geneva, Switzerland

[9]Department of Radiology, University of Southern California, Keck School of Medicine, Los Angeles, CA, USA

[10]Department of Radiology, Shohada-ye Tajrish Hospital, Shahid Beheshti University of Medical Sciences, Tehran, Iran

**Running Title:** Per-Lesion Radiomics of FAPI and FDG PET in NSCLC

## Abstract

This pilot study compares per-lesion radiomics features of [$^{68}$Ga]-DOTA FAPI-46 and [$^{18}$F]-FDG PET/CT in non-small cell lung cancer (NSCLC) to explore complementary insights into intratumoral heterogeneity beyond conventional SUV metrics, aiming to enhance lesion characterization and clinical decision-making.

A total of 28 PET/CT scans (14 [$^{18}$F]-FDG and 14 [$^{68}$Ga]-DOTA FAPI-46) were acquired for the initial staging of biopsy-confirmed NSCLC. A total of 81 co-localized lesions (lung: 21, mediastinal lymph nodes: 42, bone: 18) were segmented, with radiomics features extracted via PyRadiomics after IBSI-compliant preprocessing. Paired per-lesion comparisons used t-tests or Wilcoxon signed-rank tests with Benjamini-Hochberg FDR correction.

Significant differences (adjusted $P < 0.05$) were identified across intensity, texture, and shape features. In lung lesions, FAPI showed lower first-order metrics but higher variance and GLCM contrast, suggesting stromal heterogeneity. Mediastinal lymph nodes had fewer differences, with FAPI exhibiting lower run percentage (GLRLM: -0.298, -6.196, P=1.45E-08). Bone lesions showed extensive variations, including reduced FAPI entropy (e.g., Entropy: -1.743, -5.798, P=6.95E-08). Feature overlaps highlighted complementary stromal (FAPI) and metabolic (FDG) insights.

This pilot study demonstrates that per-lesion radiomics can capture complementary biological information from FAPI and FDG PET in NSCLC, highlighting intratumoral heterogeneity and stromal activity not fully appreciated by conventional SUV-based metrics.

## Introduction

Positron emission tomography (PET) imaging plays a central role in the management of cancer, providing crucial information for diagnosis, staging, prognostication, and therapy monitoring (1). The glucose analog [$^{18}$F]fluorodeoxyglucose ([$^{18}$F]FDG) remains the standard radiopharmaceutical in oncologic PET imaging due to its ability to visualize the increased glucose metabolism of cancer cells, a phenomenon widely known as the "Warburg effect" (2, 3). While [$^{18}$F]FDG is commonly used in clinical settings, it has notable limitations. It can show high uptake in normal tissues, such as the brain. Some tumor types, including well-differentiated neuroendocrine tumors and clear cell renal carcinoma, may take up the tracer poorly. Additionally, false-positive signals may arise during inflammation or infection, as activated immune cells also metabolize glucose at a high rate (4).

Non–small cell lung cancer (NSCLC), the most common form of lung cancer (5), poses specific challenges for imaging due to small lesions, mediastinal lymph node involvement, and variable FDG avidity across histologic subtypes (6-9). In recent years, interest has grown in exploiting the tumor microenvironment for diagnostic and therapeutic purposes. Cancer-associated fibroblasts, which support tumor growth, invasion, and immune evasion, express fibroblast activation protein (FAP) at high levels, making it an attractive imaging and therapeutic target (10-15). FAP-targeted tracers (FAP inhibitors, or FAPI) labeled with gallium-68 or fluorine-18 allow PET imaging and, via DOTA chelation, delivery of therapeutic isotopes for radioligand therapy (16-22).

FAPI PET has shown potential advantages over [$^{18}$F]FDG PET in tumor detection and delineation, particularly in lesions with low FDG uptake (23). However, most comparative studies to date have focused on semi-quantitative parameters such as SUVmax or SUVmean (24-28), and their analyses are typically performed at the per-patient level. While these parameters have clinical value, they fail to reflect the full spectrum of intratumoral heterogeneity that may drive tumor behavior and treatment response. Radiomics analysis, by extracting a broad setof quantitative descriptors related to intensity, texture, and shape, allows for a deeper assessment of tumor phenotype.

For the first time, this pilot study provides a direct per-lesion radiomics comparison of [$^{68}$Ga]-DOTA FAPI-46 and [$^{18}$F]-FDG PET in patients with NSCLC. By systematically analyzing radiomics features from these two tracers, this work explores whether they can yield complementary insights beyond conventional SUV-based metrics. Such lesion-level characterization may help capture intratumoral heterogeneity and support more precise and personalized clinical decision-making in NSCLC management.

## Material and Methods

### Patient Enrollment and Study Setting

This study was conducted at the PET/CT division of Masih-Daneshvari Hospital, affiliated with Shahid Beheshti University of Medical Sciences (SBMU). Fourteen patients with newly diagnosed, biopsy-confirmed NSCLC were included for initial staging between January 2023 and January 2024. Each patient underwent two PET/CT scans, [$^{18}$F]-FDG and [$^{68}$Ga]-DOTA FAPI-46, with an interval of 3–9 days. Written informed consent was obtained from all participants. The study was approved by the SBMU Ethics Committee (IR.SBMU.NRITLD.REC.1401.073) and registered in the Iranian Registry of Clinical Trials (IRCT: 65361). This work was financially supported by the Masih-Daneshvari PET/CT Center.

### Eligibility and Exclusion Criteria

Eligible participants were required to have: newly diagnosed NSCLC, no prior anti-cancer therapy, and the ability to undergo both PET/CT scans. Patients were excluded if they had a history of other malignancies, severe comorbidities interfering with imaging, or inability/unwillingness to provide consent.

### Radiotracer Synthesis and Quality Control

For clinical evaluation, DOTA-FAPI-46 was manually radiolabeled with gallium-68 using roughly 20 µg of the precursor and approximately 600 MBq of gallium-68. [^$^{68}$Ga]GaCl$_3$ was obtained by eluting a commercial ^$^{68}$Ge/^$^{68}$Ga generator (Pars Isotop, Iran) with 3 mL of sterile ultrapure 0.1 M HCl. A 50 µL solution of DOTA-FAPI-46 (20 µg), prepared in 500 µL of HEPES buffer, was incubated at 95 °C for 10 minutes. The reaction mixture was then purified using a Sep-Pak® C18 cartridge, removing unbound gallium-68 with two sequential water washes. The final radiotracer was eluted with 2 mL of 50% ethanol into a sterile vial and diluted with approximately 8 mL of phosphate-buffered saline (PBS). Radiolabeling efficiency and radiochemical purity (RCP) were determined using instant thin-layer chromatography (ITLC) and high-performance liquid chromatography (HPLC), respectively, yielding [^$^{68}$Ga]-FAPI-46 with a high radiochemical purity (>99%).

**PET/CT Imaging Protocol**

All scans were conducted on a GE Discovery 690 PET/CT system with Time-of-Flight (TOF) capability and a 64-slice CT. Whole-body images were acquired from the vertex to the mid-thigh. PET emission data were collected for 2–3 minutes per bed position, using a slice thickness of 3.75 mm. Low-dose CT images were obtained with slice thickness ranging from 1.33 to 2.5 mm, automatic tube current modulation (Smart mAs, 50–150 mA), a tube voltage of 120 kVp, and a helical pitch of 0.9. PET images were reconstructed using the VUE Point HD/FX algorithm, with attenuation and scatter corrections applied based on the CT data.

For [$^{18}$F]-FDG PET/CT, patients fasted for at least 6 hours, maintained blood glucose below 150 mg/dL, discontinued metformin 48–72 hours prior, and avoided strenuous activity for 24 hours before the scan. Imaging was performed 60 minutes after intravenous injection of 4.4 MBq/kg of [$^{18}$F]-FDG. In contrast, [$^{68}$Ga]-DOTA-FAPI-46 PET/CT required no special dietary or medication preparation, with scans acquired 60 minutes after intravenous administration of 2.2 MBq/kg of the tracer.

**Image segmentation**

Lesions were delineated by a radiologist with 34 years of experience and a nuclear medicine physician with over 10 years of experience using a semi-automated graph-based method (29), implemented as an extension in 3D Slicer. Lesions were categorized according to their anatomical location, including primary lung tumors, mediastinal lymph nodes, hilar lymph nodes, bone metastases, and other metastatic sites (30). Tumor segmentation was initially performed on co-registered [^$^{18}$F]-FDG PET/CT images. Following image registration (see Preprocessing section), [^$^{68}$Ga]-FAPI PET images were aligned with the FDG PET/CT scans, and lesion segmentations were refined on FAPI PET/CT images if needed. Only lesions present in both FDG and FAPI scans (i.e., co-localized tumors) were included in the comparative analysis, ensuring that paired analyses were performed on matching lesions across the two modalities. Figure 1 illustrates the workflow of this study. In Figure 2 examples of lesion segmentation on FDG PET/CT (Panel A) and the corresponding registered FAPI PET/CT (Panel B) are shown, with an example of lesion exclusion illustrated in Panel C.

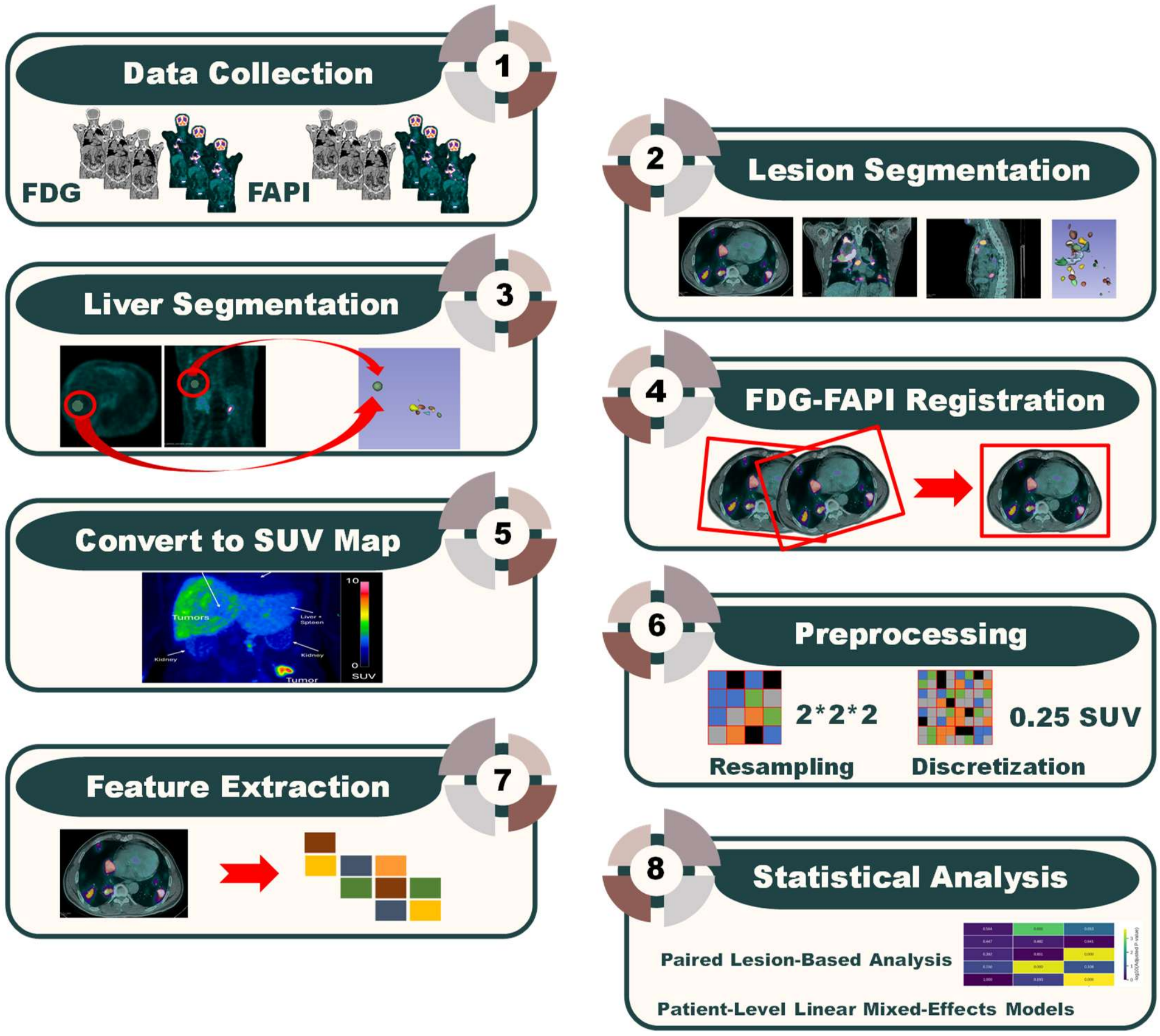

**Figure 1.** Workflow of the study.

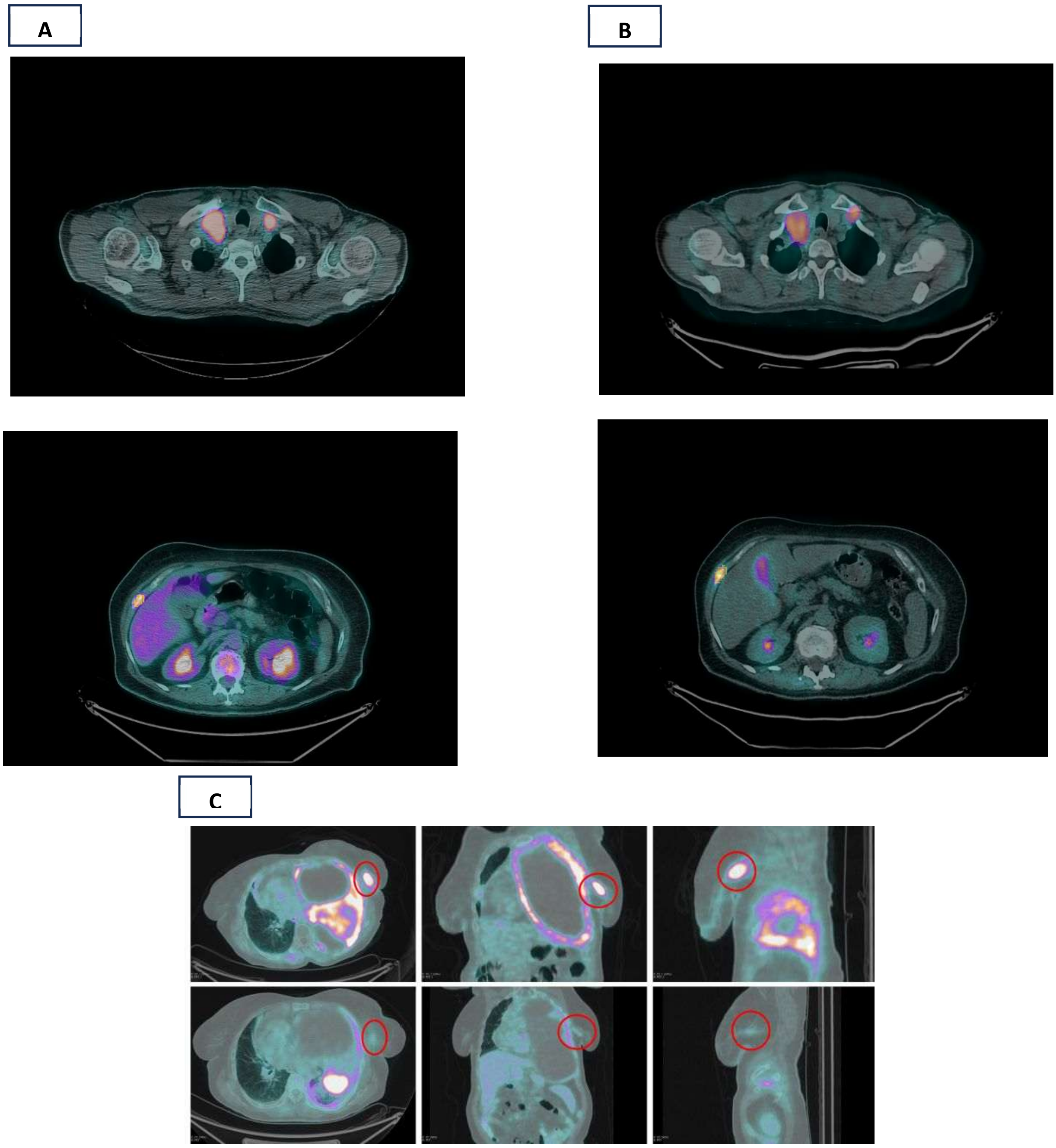

**Figure 2.** Examples of lesion segmentation and comparison between FDG and FAPI PET/CT. Panel A: Segmentation on FDG PET/CT; Panel B: Corresponding registered segmentation on FAPI PET/CT. Panel C: Top row – FAPI PET/CT showing focal uptake in the breast; bottom row – FDG PET/CT showing no uptake in the same views. Note: This image illustrates an example of lesion exclusion manually performed by the physician.

**Image Preprocessing and Radiomics Feature Extraction**

All PET/CT images were preprocessed to ensure voxel isotropy and inter-scan comparability. PET images were resampled to an isotropic voxel size of $2 \times 2 \times 2$ mm³, converted to standardized uptake value (SUV) maps, and discretized using a fixed bin size of 0.25 SUV, following the guidelines of the Image Biomarker Standardization Initiative (IBSI) (31). We used B-spline interpolation (sitkBSpline) for image resampling and nearest-neighbor interpolation (sitkNearestNeighbor) for mask resampling, with a voxel spacing of [2.0, 2.0, 2.0].

To align [68Ga]-FAPI PET/CT images with [18F]-FDG PET/CT, a multi-step registration strategy was employed using the Elastix toolkit within 3D Slicer (30). First, a rigid/affine transformations corrected global differences in rotation, translation, and scaling between the FAPI CT and FDG CT images. Subsequently, a B-spline non-rigid transformation was applied to adjust local anatomical deformations and ensure precise voxel-level alignment. The resulting transformation was then applied to the FAPI PET images, enabling accurate correspondence between the two modalities for lesion-wise analysis.

Radiomic features were extracted for each tumor using the PyRadiomics extension (32) within 3D Slicer (30). Features were extracted from SUV maps and subsequently normalized to their corresponding features from the liver VOI. Regarding clapping, no, we did not perform it.

Radiomic features were extracted separately from each tumor, and inter-patient variability was normalized using the tumor-to-liver ratio (TLR), calculated by dividing each lesion feature (e.g., SUVmax) by the corresponding feature in a spherical, lesion-free region of the liver (3 cm diameter). This normalization was applied to all features except shape, which were not adjusted using the TLR.

**Statistical Analysis**

**Paired Lesion-Based Analysis**

Paired comparisons were performed at the lesion level for tumors with matching Lesion IDs across FAPI and FDG PET/CT scans. Normality was assessed using the Shapiro-Wilk test. Normally distributed features were compared using paired t-tests, and non-normally distributed features were compared using Wilcoxon signed-rank tests. Effect sizes were calculated as Cohen’s d for t-tests and rank-biserial correlation for Wilcoxon tests. Multiple comparisons were controlled using the Benjamini-Hochberg false discovery rate (FDR) correction. Only features with ≥5 paired lesions and only tissues with ≥15 paired lesions were included in the analysis to ensure statistical robustness and reliability, and to allow valid paired comparisons between FDG and FAPI scans.

### Patient-Level Linear Mixed-Effects Models

Linear mixed-effects models (LMMs) were fitted at the patient level to account for intra-patient correlations due to multiple lesions per patient. Extended models additionally included age, sex, and tumor pathology as covariates. P-values and standardized effect sizes were reported, with FDR correction applied for multiple testing.

Significant features (adjusted $p < 0.01$ and absolute effect size $> 0.5$) were visualized using heatmaps of −log10 adjusted p-values and boxplots of lesion-level distributions. Analyses and visualizations were conducted using Python (v3.10) with the following packages: pandas (v2.2.2), numpy (v2.0.2), scipy (v1.16.2), statsmodels (v0.14.5), matplotlib (v3.10.0), and seaborn (v0.13.2).

## Results

### Patient Demographics and Lesion Characteristics

A total of 81 lesions from 14 patients with NSCLC were included in this study. The cohort comprised 9 males (64.3%) and 5 females (35.7%), with a mean age of 61 ± 6.08 years. Primary tumor pathology was predominantly adenocarcinoma (ADC, 11 patients, 78.6%) and squamous cell carcinoma (SCC, 3 patients, 21.4%). PET imaging metrics revealed a mean primary FAPI SUVmax of 9.57 ± 2.83 (median 9.95 [5.87–15.4]) and a mean primary FDG SUVmax of 12.62 ± 3.22 (median 13.13 [7.6–17.89]). Staging changes between FAPI and FDG PET occurred in a minority of patients, with downstaging in 3, upstaging in 1, and no change in 10 cases (Table 1).

**Table 1.** Summary of patient demographics, tumor characteristics, and PET imaging metrics. Values are presented as mean ± SD or counts where appropriate. SUVmax = maximum standardized uptake value; FAPI = fibroblast activation protein inhibitor; FDG = fluorodeoxyglucose; ADC = adenocarcinoma; SCC = squamous cell carcinoma.

| Feature | Value (n / % or mean ± SD / median [min–max]) |
|---|---|
| **Patients** | |
| **Total Patients** | 14 |
| **Male** | 9 (64.3%) |
| **Female** | 5 (35.7%) |
| **Age (years)** | |
| **Mean ± SD** | 61 ± 6.08 |
| **Primary FAPI SUVmax** | |
| **Mean ± SD** | 9.57 ± 2.83 |
| **Median [Min–Max]** | 9.95 [5.87–15.40] |
| **Primary FDG SUVmax** | |
| **Mean ± SD** | 12.62 ± 3.22 |
| **Median [Min–Max]** | 13.13 [7.6–17.89] |
| **Pathology** | |
| **ADC** | 11 (78.6%) |
| **SCC** | 3 (21.4%) |
| **Staging changes (FAPI vs FDG)** | |
| **Down** | 3 |
| **Up** | 1 |
| **None** | 10 |

We found metastatic lesions in several anatomical sites, such as lungs, mediastinal lymph nodes, bone, brain, liver, abdominal lymph nodes, and adrenal glands. Imaging also revealed signs of involvement in the pleura and pericardium. Detection rates differed between FAPI and FDG (Figure 3). Among 147 metastatic lesions identified, 106 (72%) were positive on both tracers, 18 (12%) were FAPI-positive/FDG-negative, and 22 (15%) were FDG-positive/FAPI-negative.

Site-specific analysis revealed that concordance was highest for pulmonary nodules (85%) and mediastinal lymph nodes (81%), whereas discordance was most pronounced in bone lesions, where 6 were FAPI-only and 10 were FDG-only. Notably, FAPI uniquely detected additional lesions in the liver and abdominal lymph nodes that were missed by FDG, while FDG detected more lesions in bone overall. These findings suggest that FAPI and FDG provide complementary information: FAPI may be more sensitive to soft-tissue and nodal involvement, while FDG captures glycolytically active skeletal metastases.

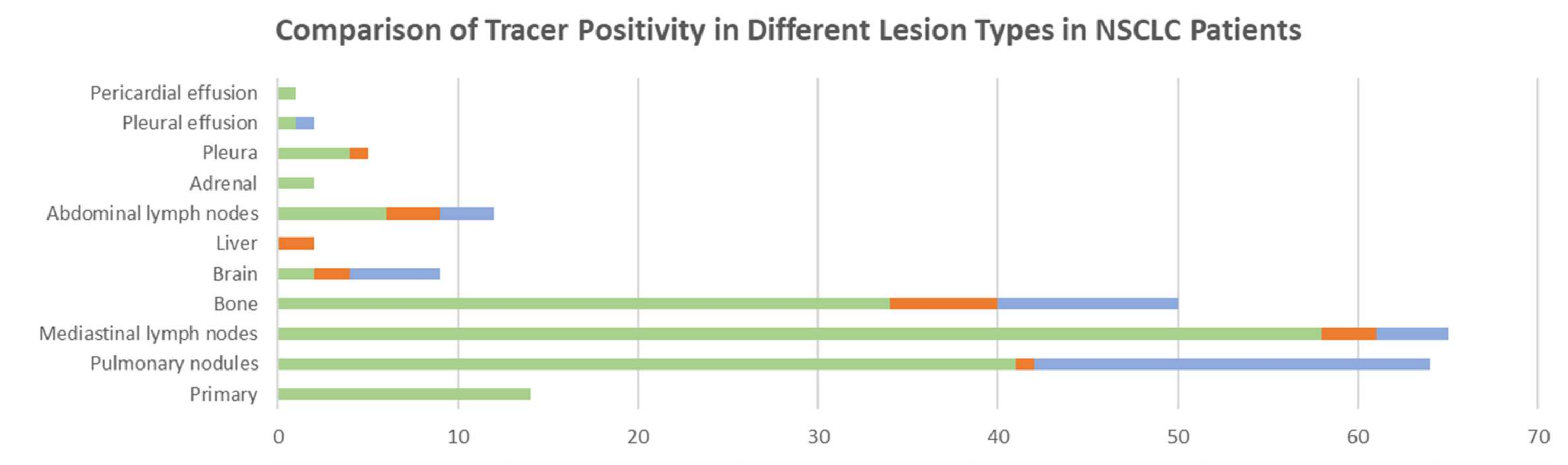

| | Primary | Pulmonary nodules | Mediastinal lymph nodes | Bone | Brain | Liver | Abdominal lymph nodes | Adrenal | Pleura | Pleural effusion | Pericardial effusion |
|---|---|---|---|---|---|---|---|---|---|---|---|
| Both positive | 14 | 41 | 58 | 34 | 2 | 0 | 6 | 2 | 4 | 1 | 1 |
| FAPI positive, FDG negative | 0 | 1 | 3 | 6 | 2 | 2 | 3 | 0 | 1 | 0 | 0 |
| FDG positive, FAPI negative | 0 | 22 | 4 | 10 | 5 | 0 | 3 | 0 | 0 | 1 | 0 |

**Figure 3.** Stacked bar chart showing the number of tracer-avid lesions across different lesion types in NSCLC patients. Colors represent tracer positivity categories: green for lesions positive in both FAPI and FDG, orange for FAPI-positive/FDG-negative lesions, and blue for FDG-positive/FAPI-negative lesions. This color coding allows quick visual comparison of tracer uptake patterns across lesion types.

For radiomics evaluation, only co-localized lesions were included. Following image co-registration of FAPI and FDG PET/CT scans, an experienced nuclear medicine physician carefully reviewed all lesions side-by-side and excluded those not present on both modalities. After this rigorous review and lesion-by-lesion matching, 21 paired lesions in the lung, 42 in mediastinal lymph nodes, and 18 in bone remained for subsequent radiomics feature extraction and analysis (Figure 4).

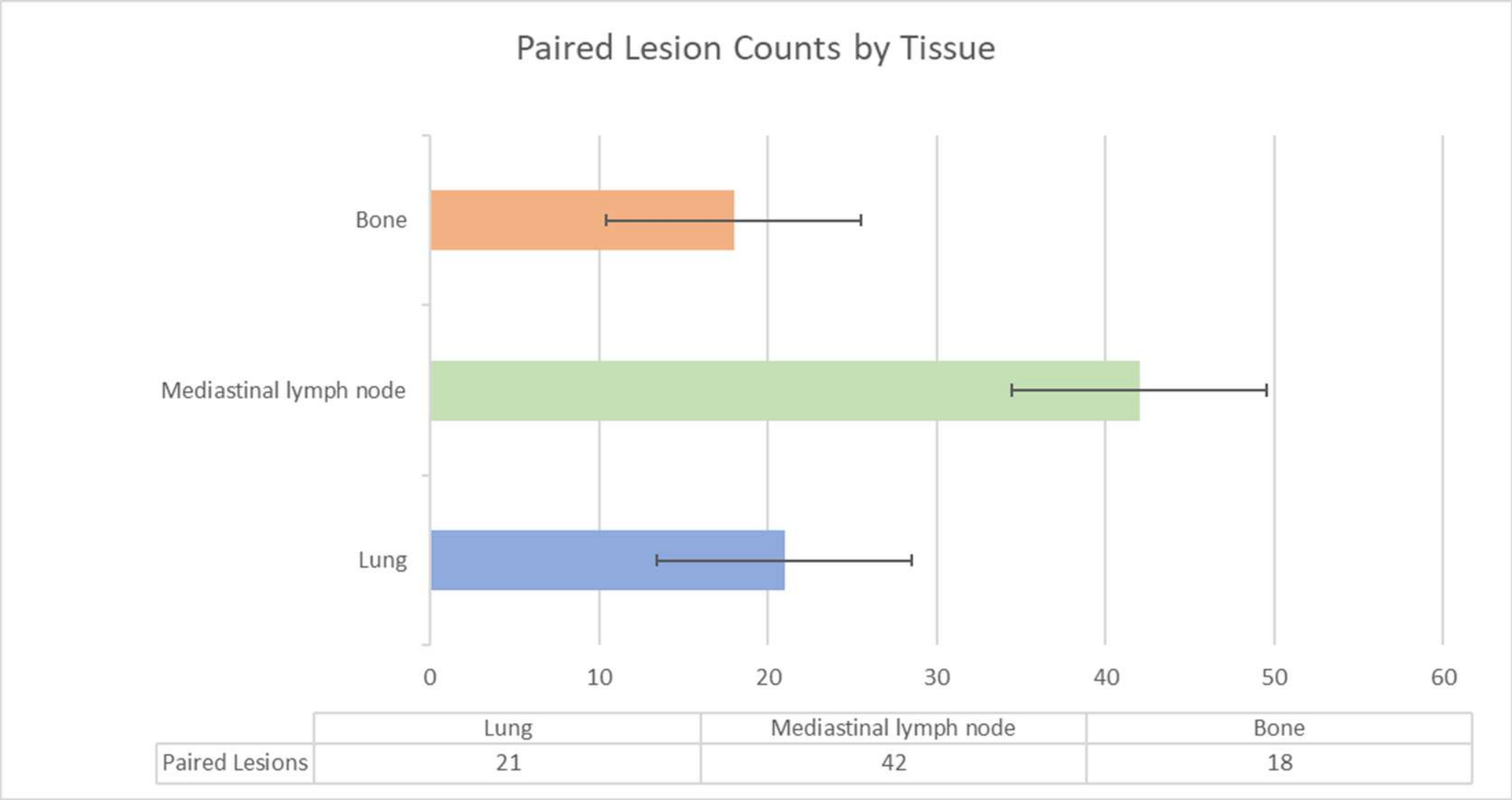

**Figure 4**. Number of paired lesions identified for each tissue type in the study. Data are shown for Lung, Mediastinal lymph node, and Bone, highlighting the availability of paired FAPI and FDG PET scans for analysis.

## Paired Per-Lesion Analysis

The full results of the paired per-lesion comparisons, including statistical tests (paired t-test or Wilcoxon signed-rank test based on normality), p-values, adjusted p-values, and effect sizes for all radiomics features across lung (n=21 lesions), mediastinal lymph node (n=42 lesions), and bone (n=18 lesions) tissues, are provided in Supplementary 01.

Figure 5 shows Venn diagrams illustrating the overlap of statistically significant radiomics features (adjusted $p < 0.05$) across lung, mediastinal lymph node, and bone tissues. Panel A corresponds to the paired per-lesion analysis (see also Supplementary 02), while Panel B represents patient-level linear mixed-effects models. The diagrams highlight both site-specific and shared features across tissues. Figure 6 shows heatmaps of effect sizes for radiomics features common to these tissues, again for paired per-lesion analysis (Figure 6A) and linear mixed-effects models (Figure 6B). Each cell represents the effect size for a specific feature in a given tissue, with P-values displayed in scientific notation, highlighting the magnitude of differences and enabling direct comparison across tissues.

Categorized boxplots of statistically significant radiomics features ($P < 0.05$) are presented for hilar lymph nodes (Figure 7), mediastinal lymph nodes (Figure 8), and lung tissue (Figure 9), with each plot showing paired feature values and corresponding test results. Complementary heatmaps of –log10(p-values) across lung, mediastinal lymph node, and bone tissues (Figures 10–15) highlight differences between FAPI and FDG PET, emphasizing modality-specific radiomics patterns across first-order and texture-based feature classes.

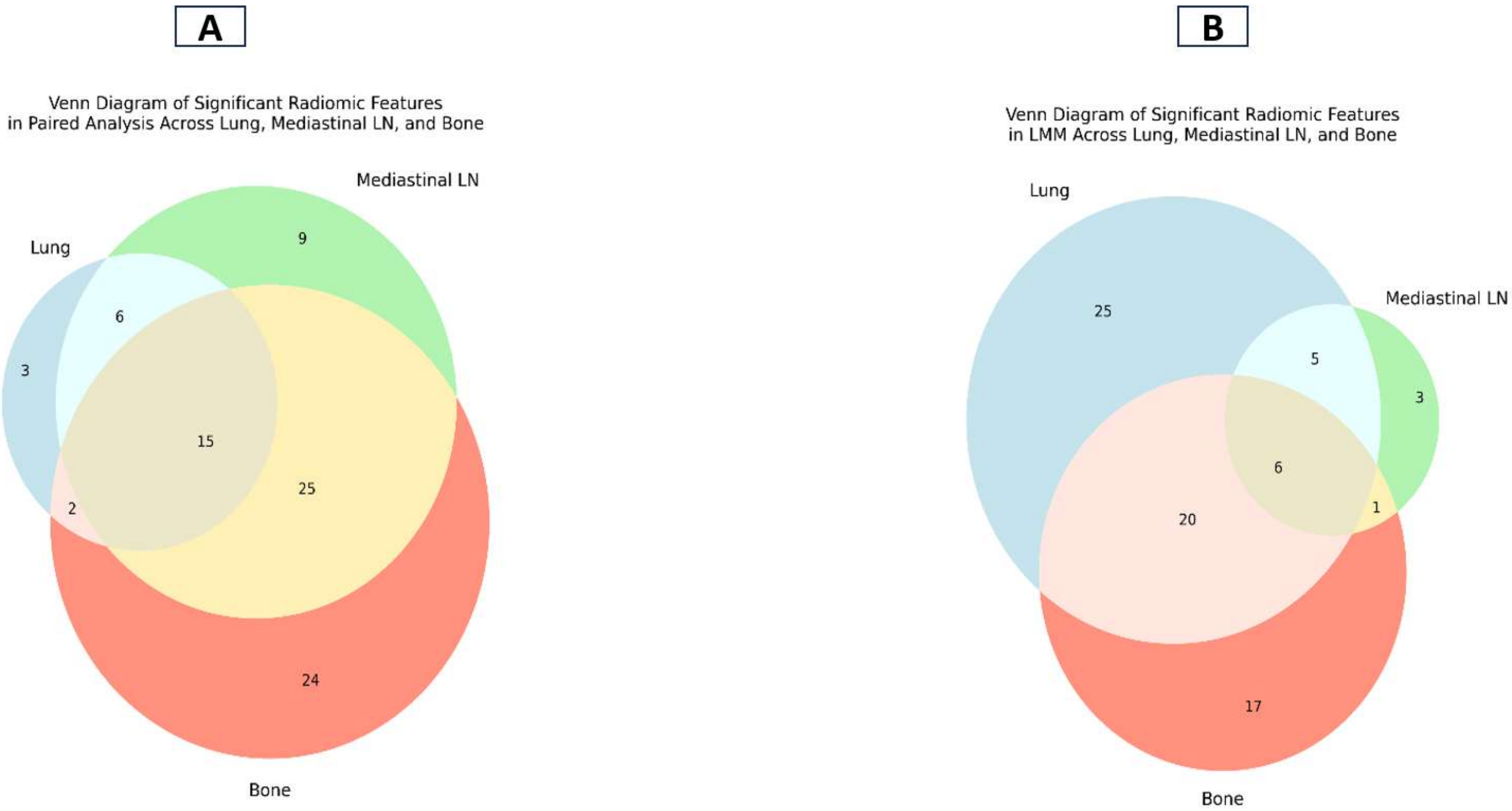

**Figure 5.** Venn diagram illustrating the overlap of significant radiomics features (Adjusted P < 0.05) among Lung, Mediastinal lymph node, and Bone in a paired per-lesion analysis. The diagram visualizes the number of features unique to each tissue as well as features shared between two or all three tissues. Colors indicate each tissue for clarity. A) Paired Lesion-Based Analysis B) Patient-Level Linear Mixed-Effects Models.

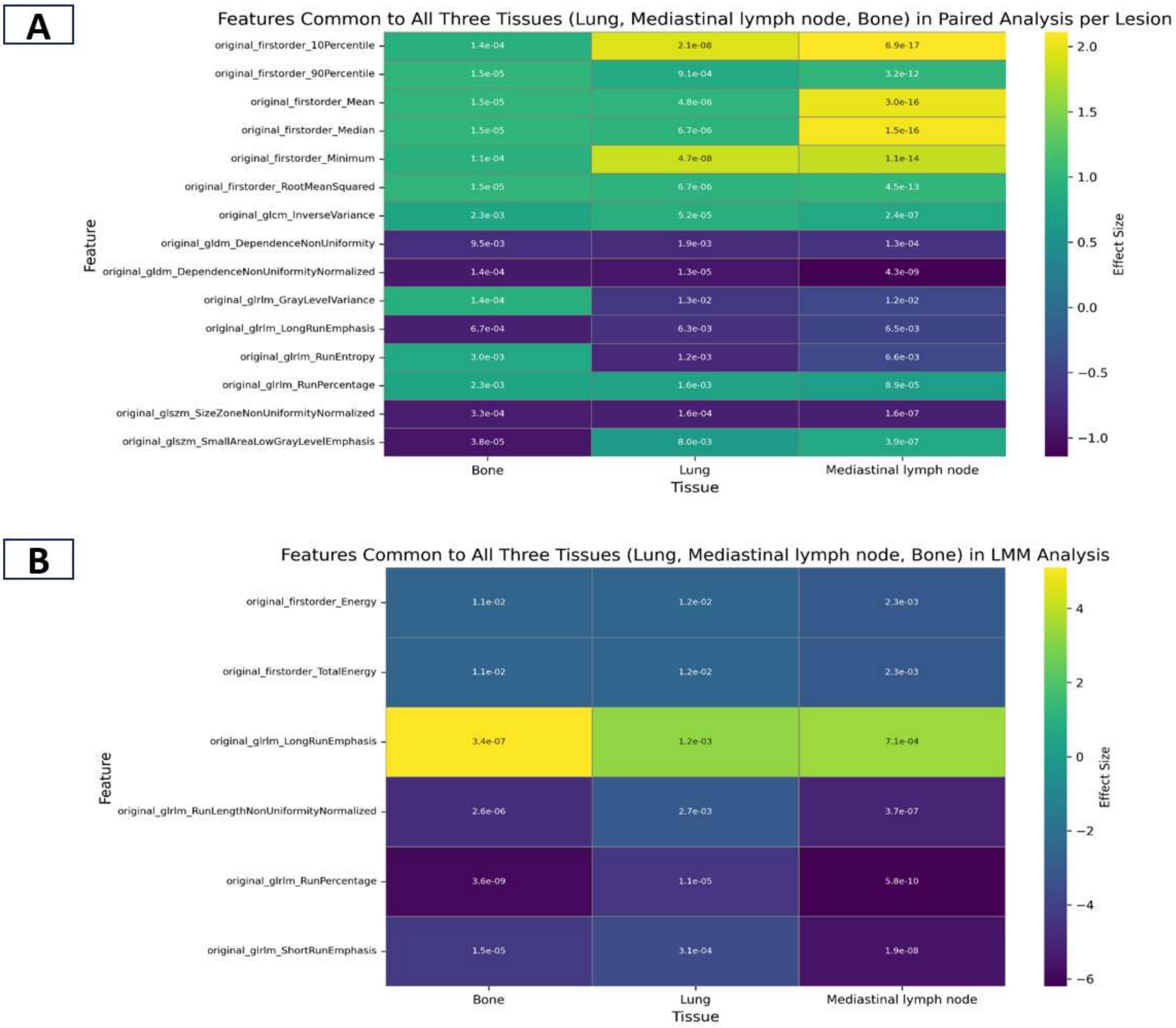

**Figure 6.** Heatmap showing the effect sizes of radiomics features common to Lung, Mediastinal lymph node, and Bone in a A) paired per-lesion analysis B) Patient-Level Linear Mixed-Effects Models. Each cell represents the effect size for a specific feature in a given tissue, with P-values displayed in scientific notation (E-format) highlights the magnitude of effect sizes, facilitating comparison across tissues.

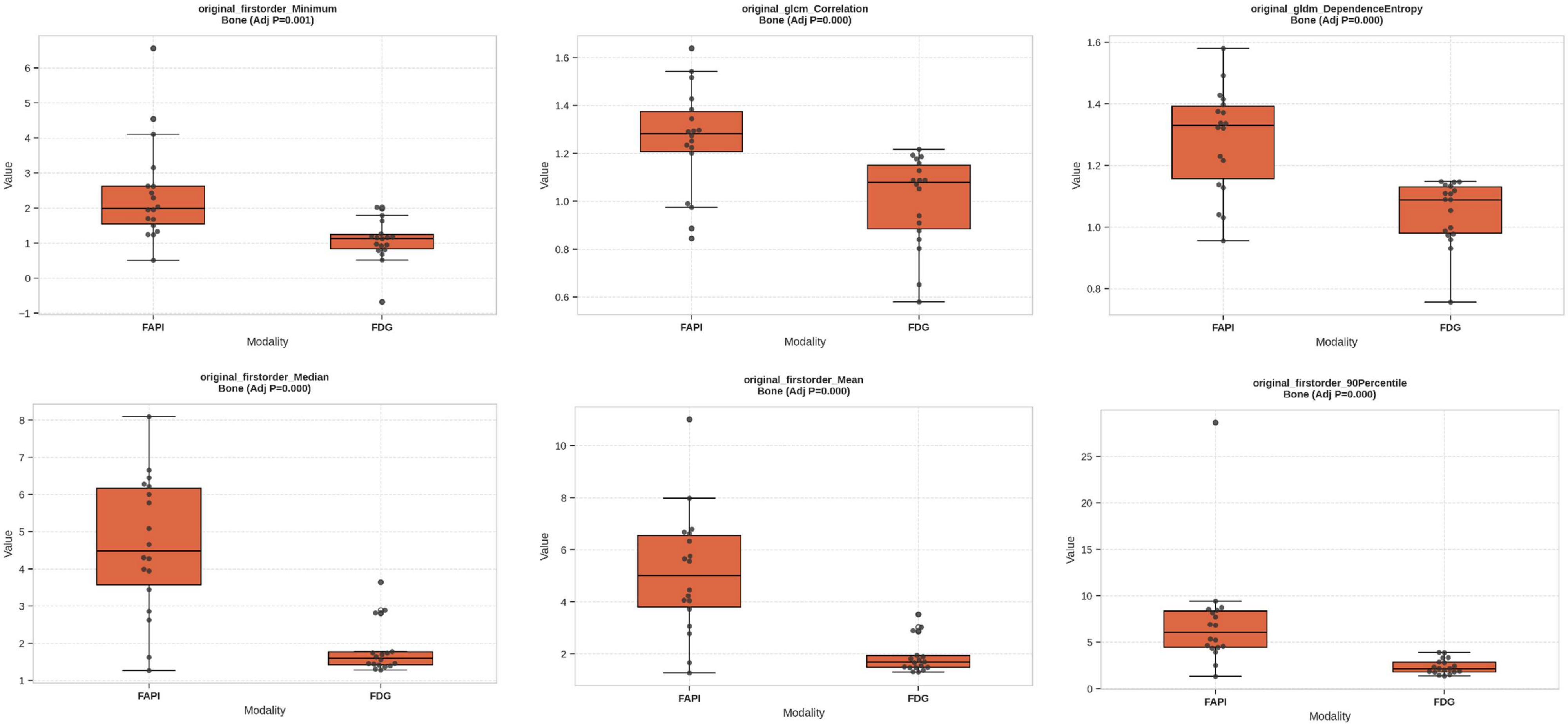

**Figure 7**. Categorized boxplots showing the distribution of radiomic features that are statistically significant (P < 0.05) between FAPI and FDG PET modalities in Hilar lymph node tissue. Each plot represents a single feature, highlighting the differences in feature values across modalities and indicating the paired P-value from the Wilcoxon signed-rank or paired t-test.

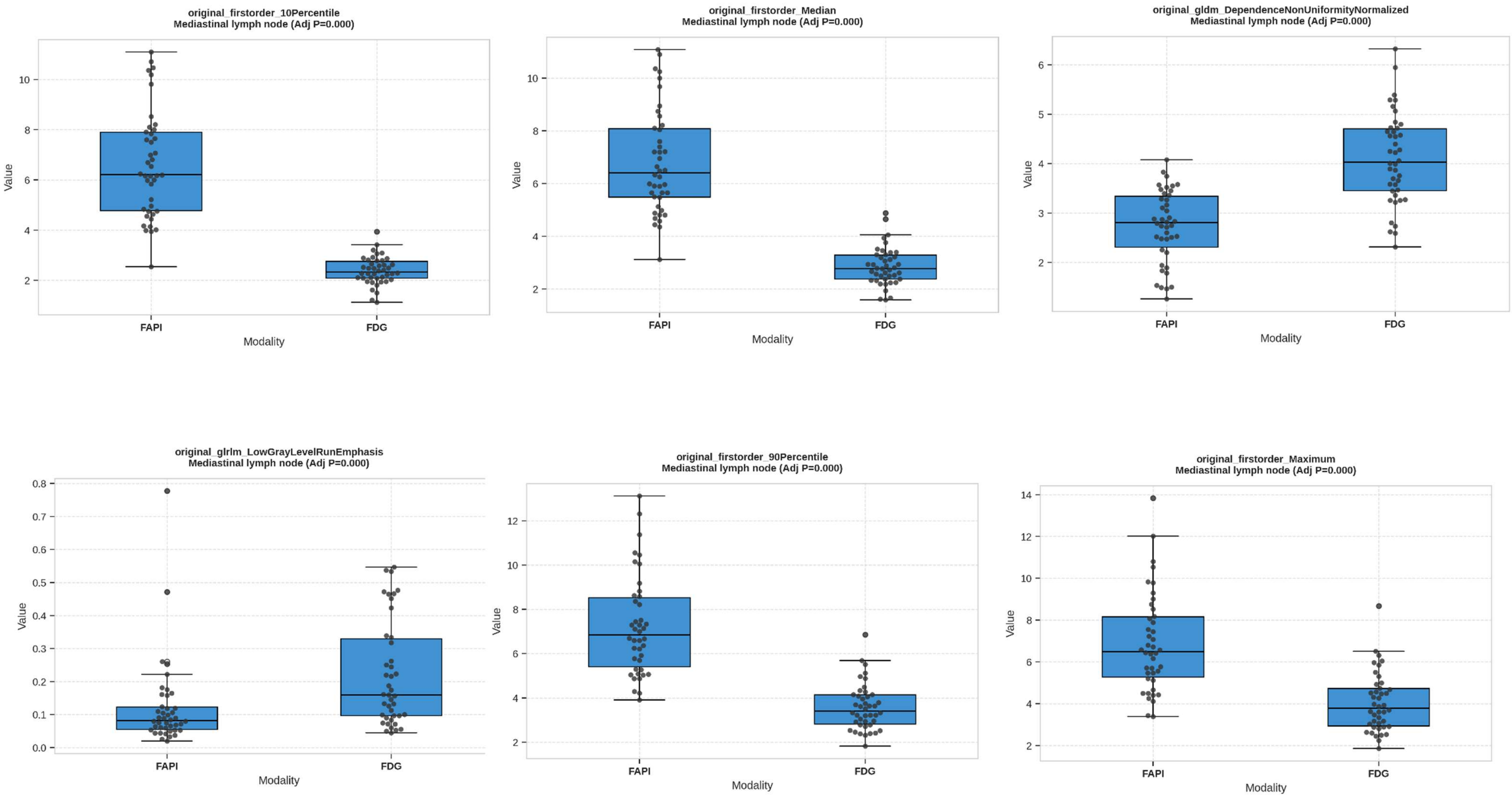

**Figure 8.** Categorized boxplots showing the distribution of radiomic features that are statistically significant ($P < 0.05$) between FAPI and FDG PET modalities in Mediastinal lymph node tissue. Each plot represents a single feature, highlighting the differences in feature values across modalities and indicating the paired P-value from the Wilcoxon signed-rank or paired t-test.

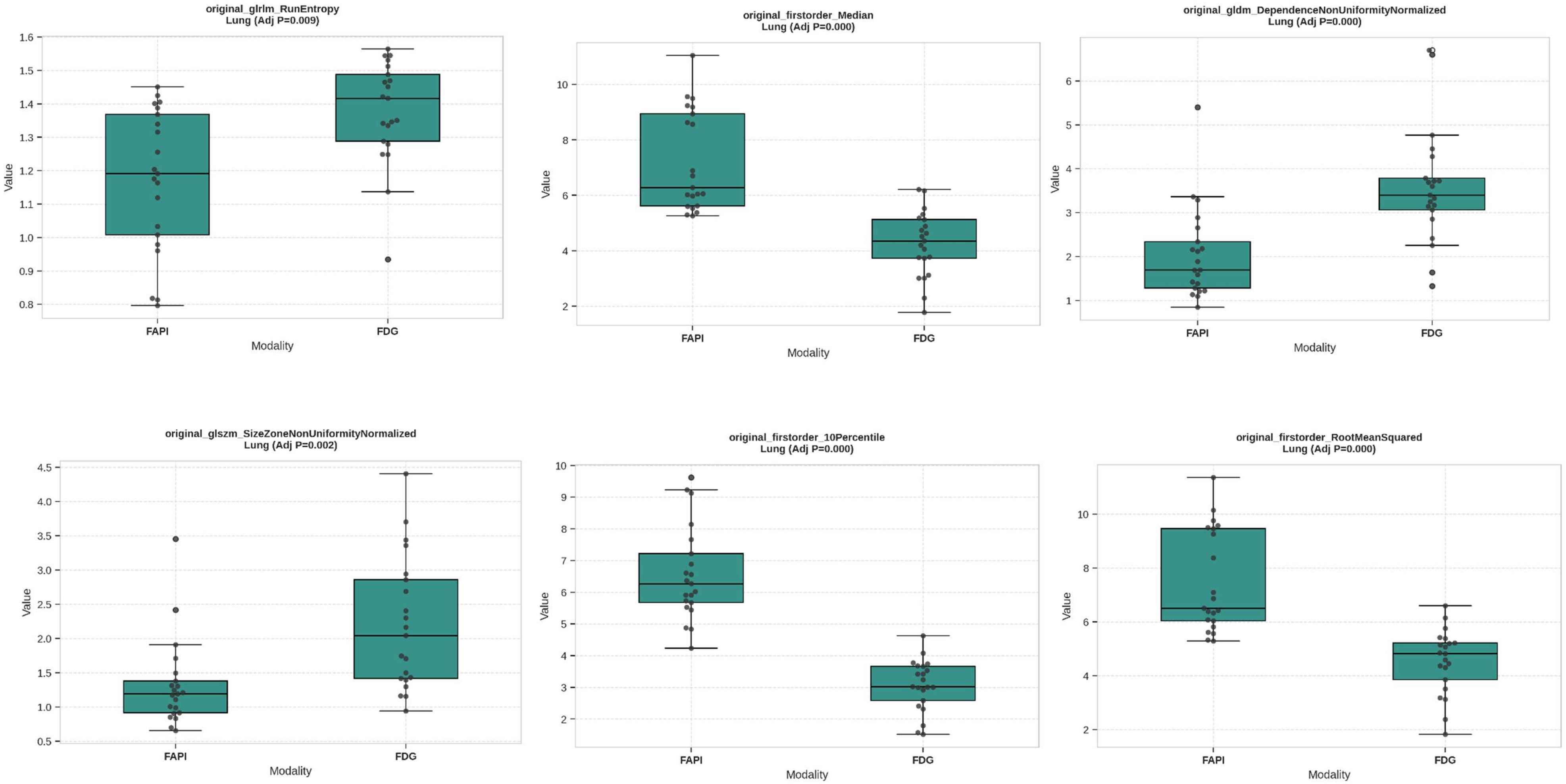

**Figure 9.** Categorized boxplots showing the distribution of radiomics features that are statistically significant (P < 0.05) between FAPI and FDG PET modalities in Lung tissue. Each plot represents a single feature, highlighting the differences in feature values across modalities and indicating the paired P-value from the Wilcoxon signed-rank or paired t-test.

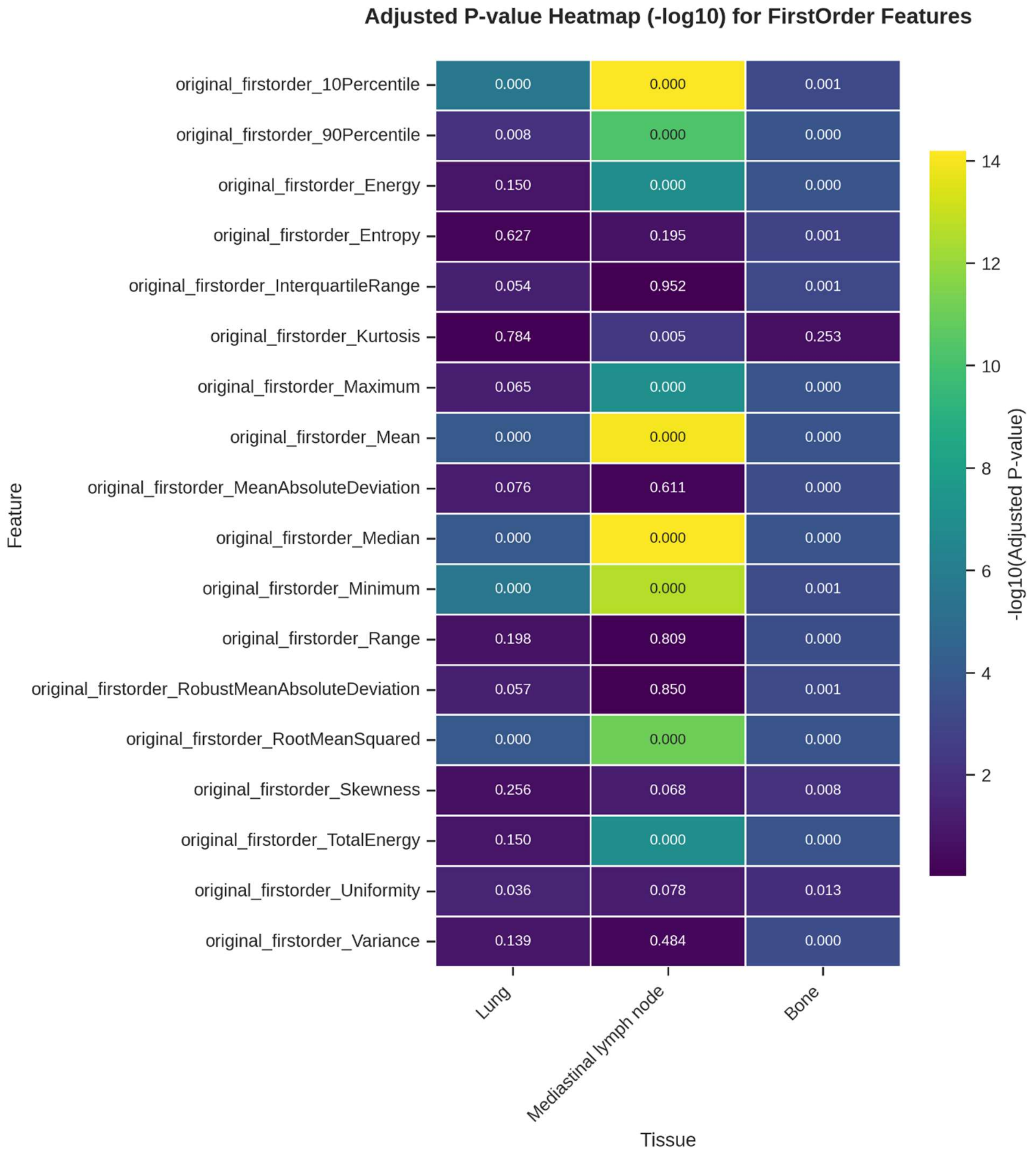

**Figure 10.** Heatmap of -log10(P-values) for first-order radiomics features across Lung, Mediastinal lymph node, and Bone, comparing FAPI and FDG PET modalities (paired t-test or Wilcoxon signed-rank test depending on normality). Significant features ($P < 0.05$) are highlighted.

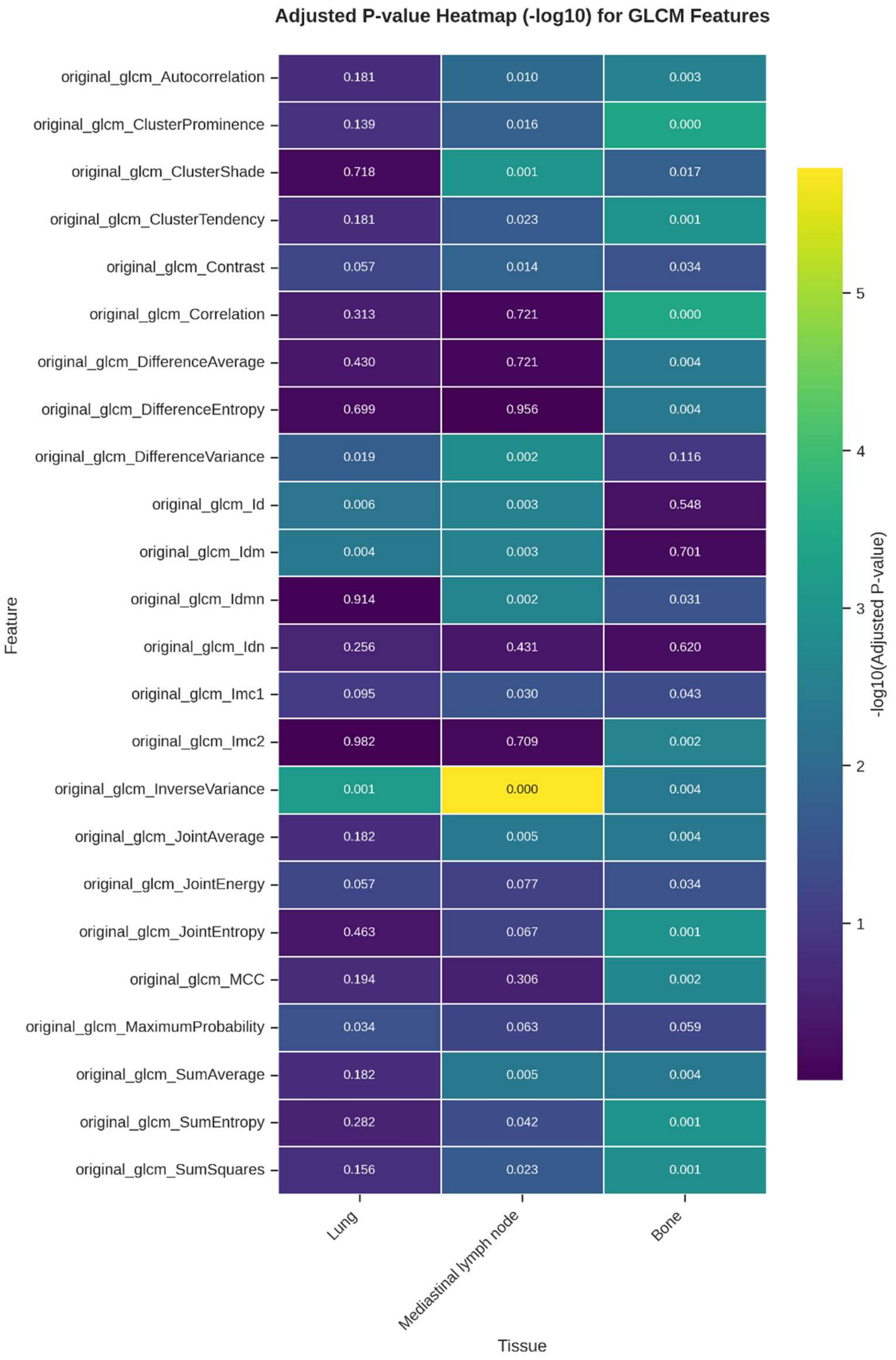

**Figure 11.** Heatmap of -log10(P-values) for GLCM features across Lung, Mediastinal lymph node, and Bone, highlighting statistically significant differences between FAPI and FDG PET modalities (paired t-test or Wilcoxon signed-rank test depending on normality). Abbreviations: GLCM – Gray-Level Co-occurrence Matrix.

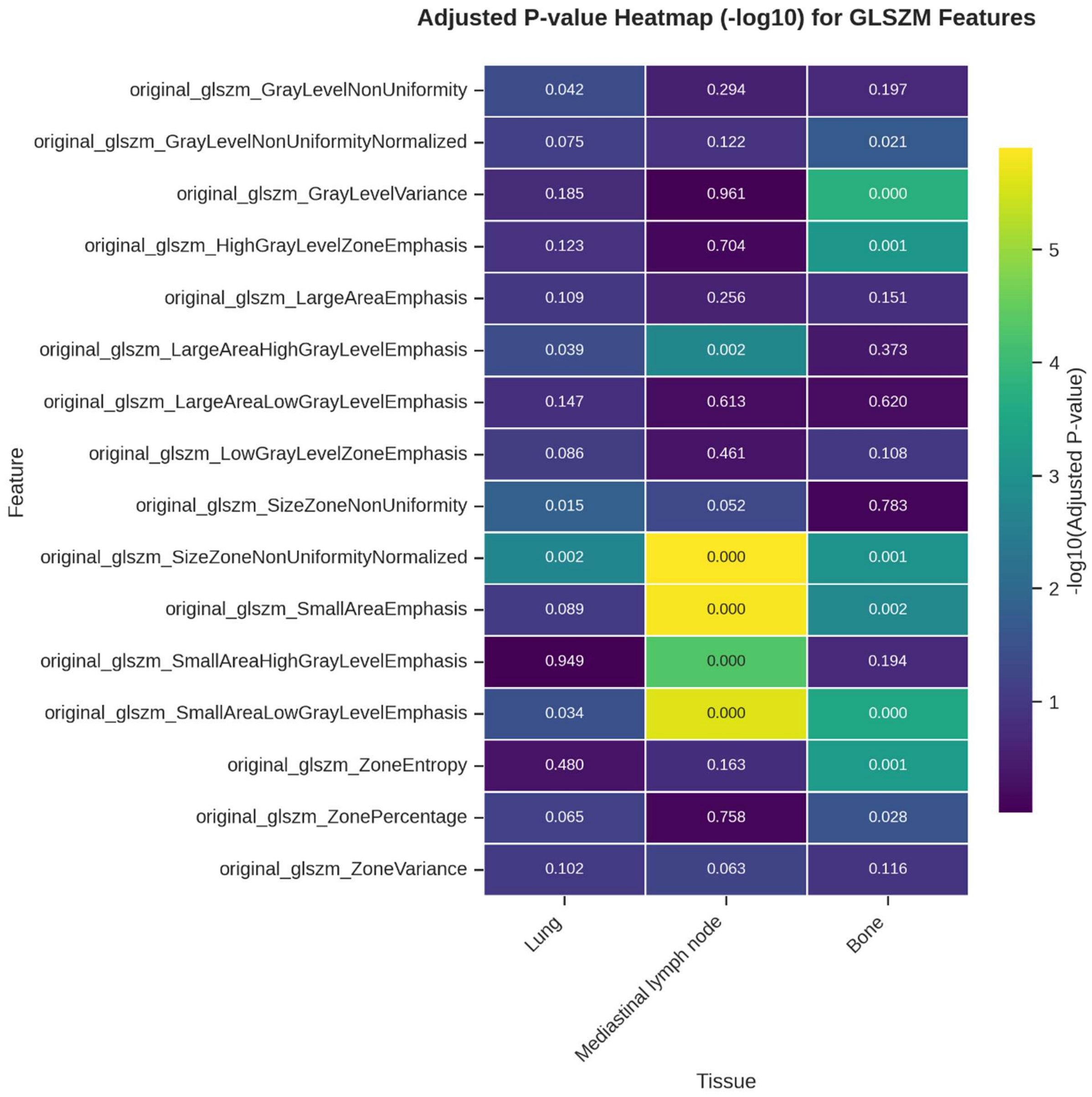

**Figure 12.** Heatmap of -log10(P-values) for GLSZM features across Lung, Mediastinal lymph node, and Bone, highlighting statistically significant differences between FAPI and FDG PET modalities (paired t-test or Wilcoxon signed-rank test depending on normality) . Abbreviations: GLSZM – Gray-Level Size Zone Matrix.

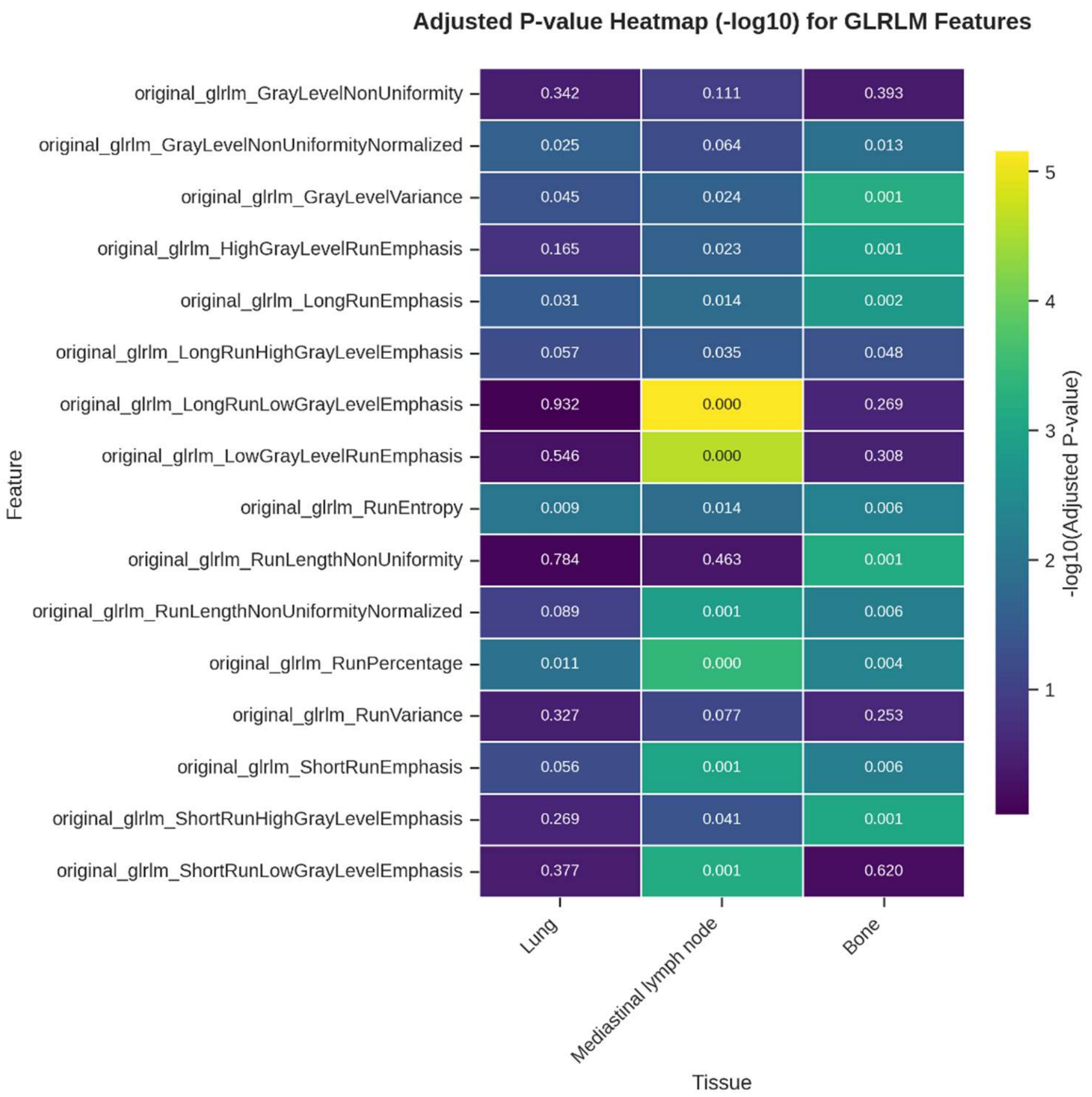

**Figure 13.** Heatmap of -log10(P-values) for GLRLM features across Lung, Mediastinal lymph node, and Bone, highlighting statistically significant differences between FAPI and FDG PET modalities (paired t-test or Wilcoxon signed-rank test depending on normality). Abbreviations: GLRLM – Gray-Level Run Length Matrix.

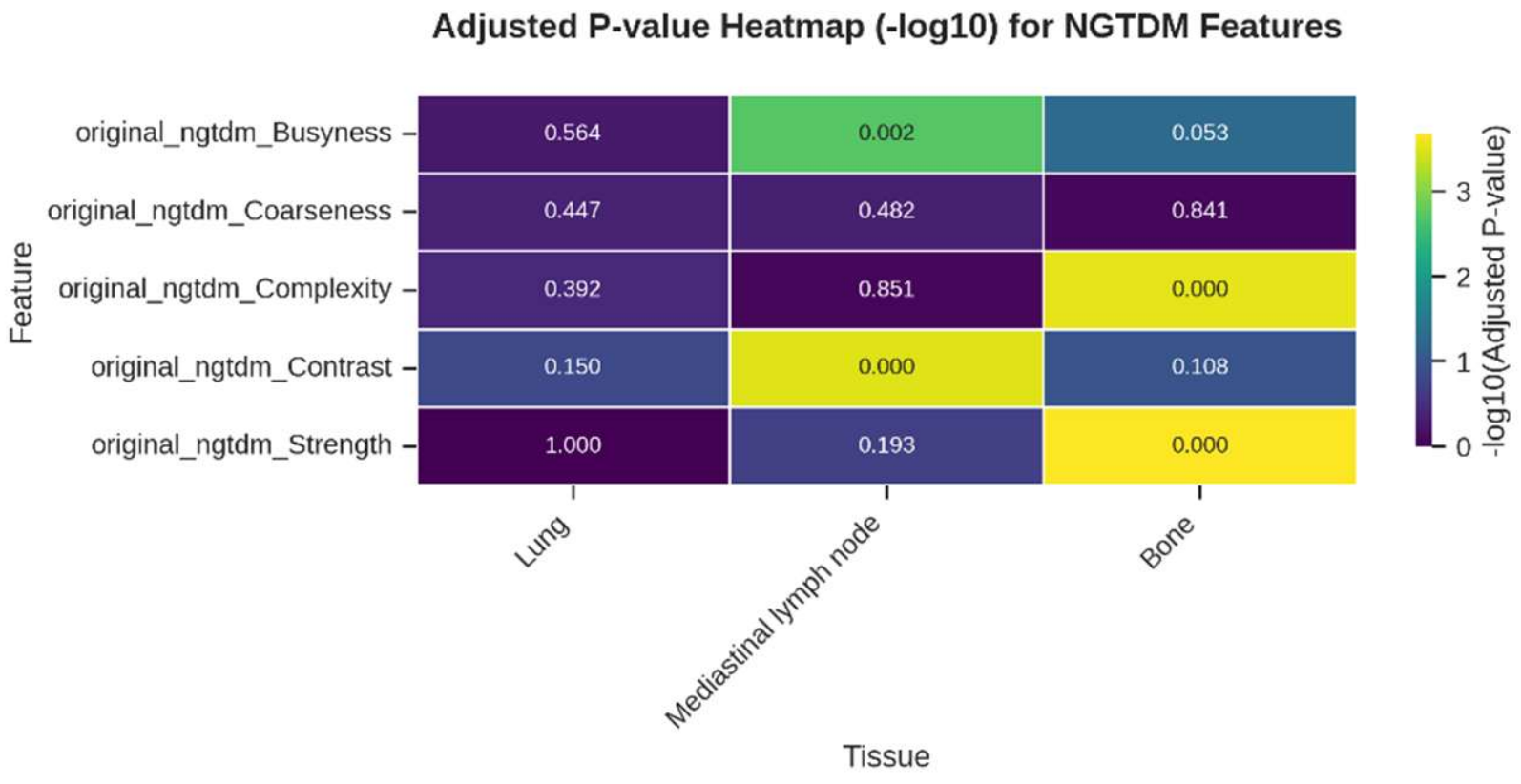

**Figure 14.** Heatmap of -log10(P-values) for NGTDM features across Lung, Mediastinal lymph node, and Bone, highlighting statistically significant differences between FAPI and FDG PET modalities (paired t-test or Wilcoxon signed-rank test depending on normality). Abbreviations: NGTDM – Neighborhood Gray Tone Difference Matrix.

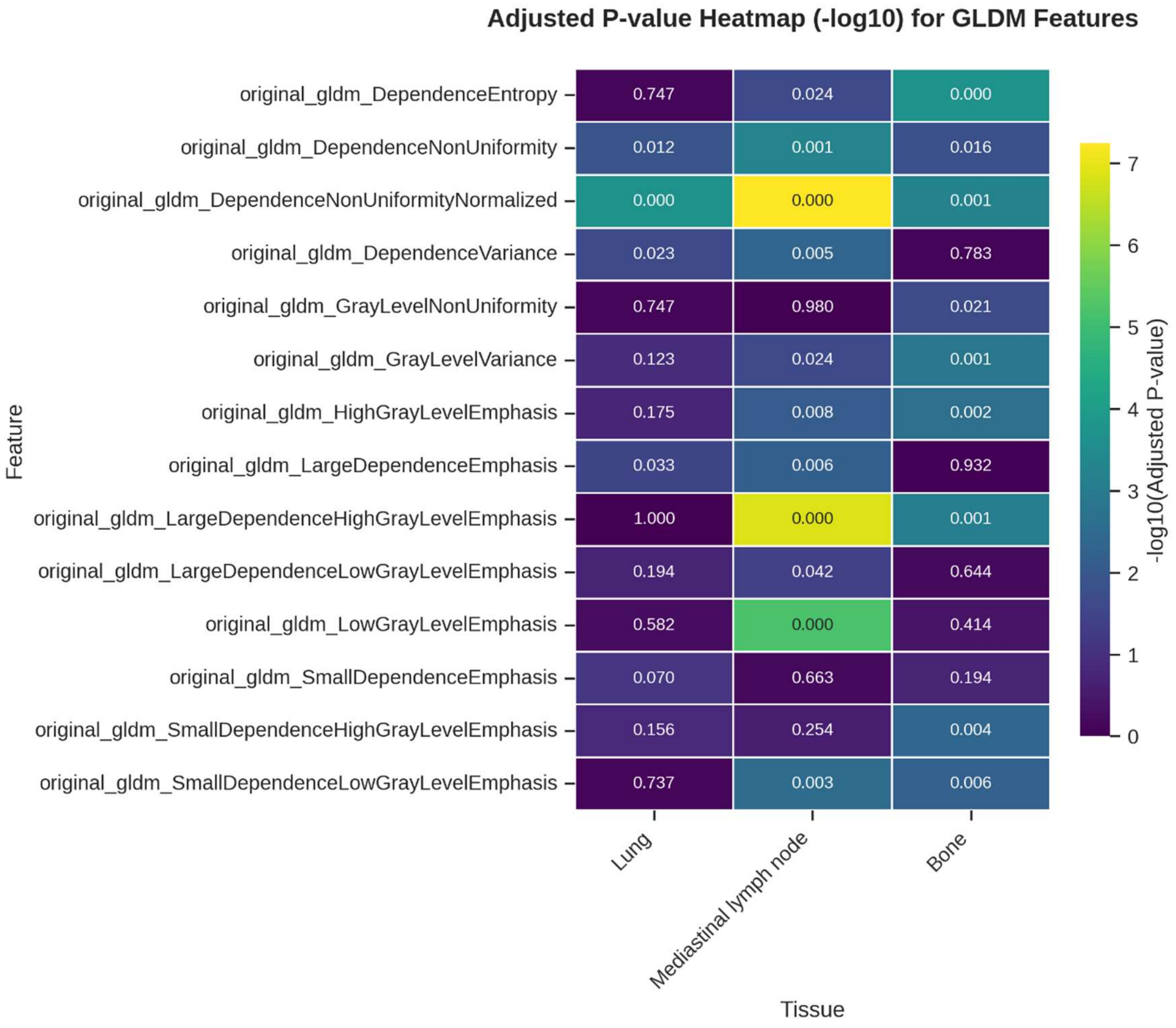

**Figure 15.** Heatmap of -log10(P-values) for GLDM features across Lung, Mediastinal lymph node, and Bone, highlighting statistically significant differences between FAPI and FDG PET modalities (paired t-test or Wilcoxon signed-rank test depending on normality). Abbreviations: GLDM – Gray-Level Dependence Matrix.

**linear mixed model (LMM) analysis**

The full linear mixed model (LMM) analysis results for all radiomics features are provided in Supplementary 03. Tables 2–4 present the significant features (adjusted P < 0.05) from this analysis, accounting for patient-level clustering. Features are reported by radiomics class, with LMM coefficients (modality effect), effect sizes, adjusted P-values, and significance levels (*P < 0.05, **P < 0.01, ***P < 0.001).

In lung tissue, 42 significant features were identified. First-order features showed strong modality effects, such as Minimum (LMM Coef: -3.267, Effect Size: -10.768, Adjusted P: 2.27E-25***), Median (-2.952, -8.522, 4.88E-16***), and 10Percentile (-3.502, -12.158, 4.83E-32***). GLCM features included Inverse Variance (-1.346, -7.432, 2.49E-12***) and Idm (-0.133, -4.718, 2.21E-05***). GLDM features like Dependence Non-Uniformity Normalized (1.538, 5.731, 1.33E-07***), GLRLM features such as Run Entropy (0.205, 4.067, 0.000341***), and GLSZM features including Size Zone Non-Uniformity Normalized (0.871, 5.680, 1.56E-07***) also exhibited notable differences (Table 2) (Supplementary 04).

For mediastinal lymph nodes, 14 significant features emerged. First-order features included Energy (-16.309, -3.052, 0.006**) and Kurtosis (0.115, 3.662, 0.001**). GLCM features such as Difference Variance (10.225, 3.433, 0.002**) and Contrast (18.300, 2.731, 0.014*), GLDM features like Large Dependence High Gray Level Emphasis (-0.787, -5.494, 3.28E-07***), and GLRLM features including Run Percentage (-0.298, -6.196, 1.45E-08***) highlighted modality-specific variations (Table 3).

In bone tissue, 39 significant features were observed. First-order features demonstrated robust effects, including Median (-2.783, -7.518, 5.17E-12***), Entropy (-1.743, -5.798, 6.95E-08***), and 10Percentile (-1.728, -5.520, 2.87E-07***). GLCM features like Sum Entropy (-1.463, -6.615, 1.15E-09***), Joint Entropy (-1.732, -6.027, 3.11E-08***), and Correlation (-0.254, -4.788, 1.05E-05***). GLDM features such as Dependence Entropy (-0.236, -7.193, 2.96E-11***), GLRLM features including Long Run Emphasis (0.042, 5.102, 2.61E-06***), and GLSZM features like Small Area Emphasis (33.345, 5.550, 2.65E-07***) underscored differences between FAPI and FDG (Table 4).

**Table 2.** Linear mixed model analysis of radiomics features in lung tissue. Features are categorized by radiomics class: Firstorder, GLCM (Gray-Level Co-occurrence Matrix), GLDM (Gray-Level Dependence Matrix), GLRLM (Gray-Level Run Length Matrix), and GLSZM (Gray-Level Size Zone Matrix). LMM Coefficients, effect sizes, and adjusted P-values are reported. Statistical significance is indicated by asterisks: P < 0.05 (*), P < 0.01 (**), P < 0.001 (***).

| Category | Feature | LMM Coef (Modality) | Effect Size | Adjusted P-value | Significance |
|---|---|---|---|---|---|
| **First order** | 10Percentile | -3.502 | -12.158 | 4.83E-32 | *** |
| **First order** | 90Percentile | -2.499 | -5.231 | 1.74E-06 | *** |
| **First order** | Energy | -172.062 | -2.525 | 0.028 | * |
| **First order** | Interquartile Range | 6.116 | 3.180 | 0.001 | ** |
| **First order** | Maximum | -2.156 | -3.067 | 0.007 | ** |
| **First order** | Mean Absolute Deviation | 5.326 | 2.940 | 0.010 | ** |
| **First order** | Mean | -2.977 | -7.045 | 3.46E-11 | *** |
| **First order** | Median | -2.952 | -8.522 | 4.88E-16 | *** |
| **First order** | Minimum | -3.267 | -10.768 | 2.27E-25 | *** |
| **First order** | Robust Mean Absolute Deviation | 6.015 | 3.153 | 0.001 | ** |
| **First order** | Root Mean Squared | -2.933 | -6.650 | 4.55E-10 | *** |
| **First order** | Total Energy | -172.062 | -2.525 | 0.028 | * |
| **First order** | Uniformity | -0.067 | -2.921 | 0.010 | ** |
| **First order** | Variance | 133.732 | 2.456 | 0.032 | * |
| **GLCM** | Cluster Prominence | 42334.345 | 2.261 | 0.048 | * |
| **GLCM** | Cluster Tendency | 92.217 | 2.349 | 0.040 | * |
| **GLCM** | Contrast | 47.270 | 3.150 | 0.001 | ** |
| **GLCM** | Difference Variance | 24.829 | 3.731 | 0.001 | ** |
| **GLCM** | Id | -0.108 | -4.498 | 5.79E-05 | *** |
| **GLCM** | Idm | -0.133 | -4.718 | 2.21E-05 | *** |
| **GLCM** | Imc1 | 0.094 | 2.357 | 0.040 | * |
| **GLCM** | Inverse Variance | -1.346 | -7.432 | 2.49E-12 | *** |
| **GLCM** | Joint Energy | -0.023 | -2.632 | 0.022 | * |
| **GLCM** | Maximum Probability | -0.036 | -3.010 | 0.008 | ** |
| **GLCM** | Sum Squares | 83.693 | 2.465 | 0.032 | * |
| **GLDM** | Dependence Non-Uniformity | 9.906 | 2.766 | 0.015 | * |
| **GLDM** | Dependence Non-Uniformity Normalized | 1.538 | 5.731 | 1.33E-07 | *** |
| **GLDM** | Dependence Variance | -0.069 | -3.099 | 0.00695 | ** |
| **GLDM** | Small Dependence Emphasis | 10.288 | 2.261 | 0.048 | * |
| **GLRLM** | Gray Level Non-Uniformity Normalized | -0.088 | -3.711 | 0.001 | ** |
| **GLRLM** | Gray Level Variance | 88.895 | 3.336 | 0.004 | ** |
| **GLRLM** | Long Run Emphasis | 0.028 | 3.246 | 0.006 | ** |
| **GLRLM** | Long Run High Gray Level Emphasis | 2.723 | 2.837 | 0.012 | * |
| **GLRLM** | Run Entropy | 0.205 | 4.067 | 0.000 | *** |
| **GLRLM** | Run Length Non-Uniformity Normalized | -0.469 | -3.006 | 0.008 | ** |
| **GLRLM** | Run Percentage | -0.351 | -4.401 | 8.37E-05 | *** |
| **GLRLM** | Short Run Emphasis | -0.232 | -3.607 | 0.002 | ** |
| **GLRLM** | Short Run Low Gray Level Emphasis | -0.244 | -2.430 | 0.033 | * |
| **GLSZM** | Gray Level Non-Uniformity | 10.849 | 3.042 | 0.008 | ** |
| **GLSZM** | Gray Level Non-Uniformity Normalized | -0.131 | -3.582 | 0.002 | ** |
| **GLSZM** | Gray Level Variance | 30.086 | 2.465 | 0.032 | * |
| **GLSZM** | High Gray Level Zone Emphasis | 21.459 | 2.852 | 0.012 | * |
| **GLSZM** | Low Gray Level Zone Emphasis | -0.172 | -3.175 | 0.006 | ** |
| **GLSZM** | Size Zone Non-Uniformity | 418.050 | 3.188 | 0.006 | ** |
| **GLSZM** | Size Zone Non-Uniformity Normalized | 0.871 | 5.680 | 1.56E-07 | *** |
| **GLSZM** | Zone Percentage | 27.568 | 3.205 | 0.006 | ** |

**Table 3.** Linear mixed model analysis of radiomics features in mediastinal lymph nodes. Features are grouped by radiomics category: Firstorder, GLCM (Gray-Level Co-occurrence Matrix), GLDM (Gray-Level Dependence Matrix), and GLRLM (Gray-Level Run Length Matrix). Reported are LMM coefficients, effect sizes, and adjusted P-values. Statistical significance is indicated as follows: P < 0.05 (*), P < 0.01 (**), P < 0.001 (***).

| Category | Feature | LMM Coef (Modality) | Effect Size | Adjusted P-value | Significance |
|---|---|---|---|---|---|
| **First order** | Energy | -16.309 | -3.052 | 0.006 | ** |
| **First order** | Kurtosis | 0.115 | 3.662 | 0.001 | ** |
| **First order** | Total Energy | -16.309 | -3.052 | 0.006 | ** |
| **GLCM** | Autocorrelation | -10.361 | -2.529 | 0.022 | * |
| **GLCM** | Contrast | 18.300 | 2.731 | 0.014 | * |
| **GLCM** | Difference Variance | 10.225 | 3.433 | 0.002 | ** |
| **GLDM** | High Gray Level Emphasis | -8.844 | -2.524 | 0.022 | * |
| **GLDM** | Large Dependence High Gray Level Emphasis | -0.787 | -5.494 | 3.28E-07 | *** |
| **GLDM** | Low Gray Level Emphasis | 0.107 | 3.722 | 0.001 | *** |
| **GLRLM** | Long Run Emphasis | 0.014 | 3.384 | 0.002 | ** |
| **GLRLM** | Long Run High Gray Level Emphasis | -0.791 | -2.246 | 0.044 | * |
| **GLRLM** | Run Length Non-Uniformity Normalized | -0.484 | -5.082 | 2.33E-06 | *** |
| **GLRLM** | Run Percentage | -0.298 | -6.196 | 1.45E-08 | *** |
| **GLRLM** | Short Run Emphasis | -0.214 | -5.618 | 2.41E-07 | *** |

**Table 4.** Linear mixed model analysis of radiomics features in bone tissue. Features are categorized into Firstorder, GLCM (Gray-Level Co-occurrence Matrix), GLDM (Gray-Level Dependence Matrix), GLRLM (Gray-Level Run Length Matrix), and GLSZM (Gray-Level Size Zone Matrix). Shown are LMM coefficients, effect sizes, and adjusted P-values. Statistical significance is indicated as follows: P < 0.05 (*), P < 0.01 (**), P < 0.001 (***).

| Category | Feature | LMM Coef (Modality) | Effect Size | Adjusted P-value | Significance |
|---|---|---|---|---|---|
| **First order** | 10 Percentile | -1.728 | -5.520 | 2.87E-07 | *** |
| **first order** | 90 Percentile | -4.829 | -3.550 | 0.002 | ** |
| **First order** | Energy | -12.687 | -2.536 | 0.028 | * |
| **first order** | Entropy | -1.743 | -5.798 | 6.95E-08 | *** |
| **first order** | Interquartile Range | -10.388 | -2.926 | 0.012 | * |
| **first order** | Maximum | -5.295 | -2.829 | 0.014 | * |
| **first order** | Mean | -3.173 | -6.043 | 3.11E-08 | *** |
| **first order** | Median | -2.783 | -7.518 | 5.17E-12 | *** |
| **first order** | Minimum | -1.334 | -4.854 | 8.66E-06 | *** |
| **first order** | Robust Mean Absolute Deviation | -10.931 | -2.599 | 0.026 | * |
| **first order** | Root Mean Squared | -3.684 | -4.792 | 1.05E-05 | *** |
| **first order** | Skewness | -4.585 | -3.355 | 0.003 | ** |
| **first order** | Total Energy | -12.687 | -2.536 | 0.028 | * |
| **GLCM** | Correlation | -0.254 | -4.788 | 1.05E-05 | *** |
| **GLCM** | Difference Average | -7.413 | -2.876 | 0.013 | * |
| **GLCM** | Difference Entropy | -0.990 | -3.350 | 0.003 | ** |
| **GLCM** | Idmn | -0.006 | -2.539 | 0.028 | * |
| **GLCM** | Imc1 | -0.231 | -2.433 | 0.037 | * |
| **GLCM** | Imc2 | -0.289 | -5.858 | 5.54E-08 | *** |
| **GLCM** | Inverse Variance | -0.815 | -3.866 | 0.001 | *** |
| **GLCM** | Joint Average | -1.437 | -2.390 | 0.039 | * |
| **GLCM** | Joint Entropy | -1.732 | -6.027 | 3.11E-08 | *** |
| **GLCM** | MCC | -0.223 | -4.436 | 5.02E-05 | *** |
| **GLCM** | Sum Average | -1.437 | -2.390 | 0.039 | * |
| **GLCM** | Sum Entropy | -1.463 | -6.615 | 1.15E-09 | *** |
| **GLDM** | Dependence Entropy | -0.236 | -7.193 | 2.96E-11 | *** |
| **GLDM** | Dependence Non-Uniformity | 0.439 | 2.843 | 0.014 | * |
| **GLDM** | Dependence Non-Uniformity Normalized | 1.362 | 5.855 | 5.54E-08 | *** |
| **GLDM** | Gray Level Non-Uniformity | 0.021 | 2.775 | 0.016 | * |
| **GLDM** | Large Dependence High Gray Level Emphasis | -0.455 | -2.770 | 0.016 | * |
| **GLDM** | Small Dependence Low Gray Level Emphasis | 3.428 | 2.909 | 0.012 | * |
| **GLRLM** | Long Run Emphasis | 0.042 | 5.102 | 2.61E-06 | *** |
| **GLRLM** | Run Entropy | -0.092 | -2.327 | 0.045 | * |
| **GLRLM** | Run Length Non-Uniformity | -1.581 | -3.157 | 0.006 | ** |
| **GLRLM** | Run Length Non-Uniformity Normalized | -0.778 | -4.699 | 1.52E-05 | *** |
| **GLRLM** | Run Percentage | -0.564 | -5.900 | 5.54E-08 | *** |
| **GLRLM** | Short Run Emphasis | -0.298 | -4.333 | 7.60E-05 | *** |
| **GLSZM** | Size Zone Non-Uniformity Normalized | 0.652 | 3.580 | 0.001 | ** |
| **GLSZM** | Small Area Emphasis | 33.345 | 5.550 | 2.65E-07 | *** |
| **GLSZM** | Small Area Low Gray Level Emphasis | 5.697 | 3.354 | 0.003 | ** |
| **GLSZM** | Zone Entropy | -0.367 | -3.624 | 0.001 | ** |

## Discussion

To our knowledge, this pilot study represents the first direct per-lesion radiomics comparison between [$^{68}$Ga]-DOTA FAPI-46 and [$^{18}$F]-FDG PET imaging in patients with newly diagnosed NSCLC, focusing on quantitative features that extend beyond traditional SUV metrics. By analyzing intensity, texture, and shape-based radiomics in paired lesions across lung, mediastinal lymph nodes, and bone tissues, we identified significant modality-specific differences, offering complementary biological insights into tumor metabolism and the stromal microenvironment. These findings suggest that multimodal radiomics can better capture intratumoral heterogeneity and may ultimately support more precise staging, prognostication, and therapeutic strategies in NSCLC.

### Key Radiomics Differences and Their Biological Correlates

In primary lung lesions, the linear mixed model (LMM) showed significant differences in several radiomics features between FAPI and FDG PET imaging. Among first-order features, parameters such as the Minimum, Median, and 10th Percentile were consistently lower in FAPI compared to FDG (Table 2). This pattern suggests reduced tracer intensity in stromal-targeted imaging, reflecting the distinct uptake behavior of FAPI in tumor microenvironments. These results are consistent with a prior report (33) showing higher SUVmax for FAPI tracers such as 68Ga-FAP-2286 (9.90 ± 5.61) compared with FDG (6.09 ± 2.84, $P < 0.01$), while maintaining a 100% detection sensitivity. Paired analyses (Supplementary 01) confirmed that FAPI exhibits consistently lower intensity-related feature values than FDG, underscoring its distinct imaging profile. Biologically, this pattern may reflect the uneven distribution of cancer-associated fibroblasts (CAFs). These cells tend to cluster in the peritumoral region rather than the glycolysis-dominant tumor core. FAP-targeted tracers such as FAPI show minimal uptake in normal tissues but strong expression in most epithelial cancers (over 90%) (34). In contrast, FAPI showed higher values for variance-related features, including the Interquartile Range ($P = 0.001$) and Variance ($P = 0.014$) (Table 2). These increases indicate greater stromal heterogeneity, potentially driven by hypoxia and TGF-β signaling. (35). The paired analysis supported this observation. The Wilcoxon signed-rank test for the Interquartile Range showed a significant trend toward greater heterogeneity in FAPI ($P = 0.016$) (Supplementary 01). Additional comparisons revealed that FAPI exhibits a lower tumor-to-background ratio (TBR) in primary tumors (25.3 ± 14.0 vs. 32.1

± 21.1 for FDG, $P < 0.001$). This finding emphasizes that FAPI preferentially highlights stromal regions rather than overall metabolic activity (27, 33).

Texture-based features derived from the gray-level co-occurrence matrix (GLCM) and gray-level run-length matrix (GLRLM) were generally higher in FAPI (e.g., Contrast, $P = 0.002$; Inverse Variance, $P < 0.001$; Run Entropy, $P < 0.001$) (Table 2). This indicates coarser and more heterogeneous uptake, consistent with the fibrotic tumor stroma (36). Paired analyses reinforced these observations (Supplementary 1). GLCM Contrast showed a Wilcoxon statistic of 49 ($P = 0.020$), and GLRLM Run Entropy a statistic of 27 ($P = 0.001$), indicating increased heterogeneity in FAPI. Additional GLRLM features, including Long Run Emphasis ($P = 0.001$) and Short Run Emphasis ($P < 0.001$), further highlighted irregular stromal patterns.

Gray-level size zone matrix (GLSZM) features, such as Small Area Emphasis ($P = 0.047$) and Zone Percentage ($P = 0.001$), suggested the presence of localized stromal compartments. Paired tests confirmed this, with GLSZM Small Area Emphasis showing a Wilcoxon statistic of 56 ($P = 0.038$), consistent with patchy CAF distributions.

These textural differences likely reflect FAPI's advantage in visualizing stromal fibrosis compared with FDG's glycolytic focus. This property may enhance primary tumor delineation in non-small cell lung cancer (NSCLC).(23).

In mediastinal lymph nodes, fewer but meaningful differences were observed in the LMM analysis (Table 3). GLRLM features, including Run Percentage ($P < 1\times10^{-9}$) and Run Length Non-Uniformity Normalized ($P < 1\times10^{-6}$), as well as GLDM descriptors such as Large Dependence High Gray Level Emphasis ($P < 1\times10^{-7}$), were generally lower in FAPI.

In contrast, FDG showed higher first-order Energy ($P = 0.002$) and Kurtosis ($P < 0.001$), consistent with glycolytic activity likely driven by immune cell infiltration. These findings suggest that FAPI highlights stromal features, whereas FDG primarily reflects metabolic activity in immune-rich regions. (37). Paired analyses in Supplementary 01 (42 lesions) confirmed these observations. GLRLM Run Percentage showed a Wilcoxon statistic of 152 ($P < 0.001$), and GLDM Large Dependence High Gray Level Emphasis a statistic of 51 ($P < 1\times10^{-8}$), indicating more clustered uptake patterns in FAPI.

The reduced dependence emphasis in FAPI likely reflects FAP expression primarily on perinodal fibroblasts, which tend to form clustered, less diffuse patterns compared with the widespread glycolytic activity captured by FDG. This is supported by higher SUVmax in FAPI lymph nodes (7.95 ± 2.75 vs. 5.55 ± 1.59 for FDG, P = 0.01) and 100% sensitivity versus 78.8% for FDG, suggesting improved distinction of metastatic from reactive nodes through CAF targeting.

These differences may also explain the higher contrast and Difference Variance in FAPI GLCM features (Difference Variance P = 0.001; Contrast P = 0.006), which likely reflect stromal barriers that restrict lymphatic spread and contribute to nodal fibrosis. Paired tests further supported elevated GLCM Difference Variance (Wilcoxon statistic: 180, P < 0.001) and Inverse Variance (Wilcoxon statistic: 76, $P < 1\times10^{-6}$), highlighting the textural disparities linked to fibrotic remodeling.

Furthermore, FAPI demonstrated superior tumor-to-background ratio (TBR) in metastatic lymph nodes (7.5 ± 6.6 vs. 5.9 ± 8.6 for FDG, P < 0.001) and higher specificity for nodes smaller than 10 mm (100% vs. 17%, P < 0.001), enhancing staging accuracy. (33).

In bone metastases, the largest number of significant differences was observed in the LMM analysis (36 observations, Table 4). First-order features such as Entropy and Skewness were lower in FAPI, suggesting more symmetric and less variable uptake. In contrast, FDG captured glycolytic bursts from osteoclastic activity, reflected by higher intensity percentiles, including the 10th and 90th Percentiles.

Paired analyses in Supplementary 01 (18 lesions) confirmed these findings. Entropy showed a Wilcoxon statistic of 12 (P = 0.001) and Skewness a statistic of 23 (P = 0.005), reinforcing the pattern of more uniform uptake in FAPI.

Texture features, including GLCM entropy metrics such as Joint Entropy and Sum Entropy, were also reduced in FAPI. This indicates more homogeneous stromal infiltration compared with the heterogeneous metabolic activity observed with FDG. Interestingly, FAPI showed elevated GLSZM Small Area Emphasis (LMM P-value: 2.85E-08, Coefficient: 33.345, Effect Size: 5.550, Adjusted P-value: 2.65E-07) and Size Zone Non-Uniformity Normalized (LMM P-value: 0.000, Coefficient: 0.652, Effect Size: 3.580, Adjusted P-value: 0.001), possibly linked to localized niches of FAP-expres (38)sing fibroblasts within osteoid tissue (39, 40). Paired tests validated these observations. Small Area Emphasis showed a Wilcoxon statistic of 13 ($P = 0.001$, Effect Size = -0.848, Adjusted $P = 0.002$), and Small Area Low Gray Level Emphasis a statistic of 3 ($P < 4\times10^{-5}$, Effect Size = -0.965, Adjusted $P = 0.000$).

GLRLM features such as Long Run Emphasis ($P < 4\times10^{-7}$) and Run Percentage ($P < 4\times10^{-9}$) further indicated structured stromal patterns in FAPI. Paired Wilcoxon tests supported these results (Long Run Emphasis: statistic 13, $P = 0.001$; Run Percentage: statistic 19, $P = 0.002$), highlighting FAPI's ability to capture organized stromal architecture.

These differences suggest that FAPI emphasizes fibrotic microenvironments in bone lesions, whereas FDG mainly reflects metabolic heterogeneity. FAPI showed higher SUVmax (7.74 ± 3.72 vs. 5.66 ± 3.55; $P = 0.04$) and 100% sensitivity compared with 68.5% for FDG, particularly in skull and rib metastases where physiological uptake is minimal.

FAPI also exhibited a higher tumor-to-background ratio in bone metastases (8.6 ± 5.4 vs. 4.3 ± 2.3; $P < 0.001$) and superior sensitivity (99% vs. 89%; $P = 0.002$), likely due to FAP expression in osteoblasts and osteocytes, key components of bone stroma. Overall, these radiomics findings suggest that FAPI PET could complement or even surpass FDG in oncology staging. Meta-analyses further support this, showing higher sensitivity (0.96 vs. 0.73) and specificity (0.92 vs. 0.83) for cancer diagnosis (33).

**Radiomics Feature Overlap and Dual-Tracer Insights**

Venn diagram analyses from both the per-lesion (Figure 5A) and patient-level LMM (Figure 5B) approaches revealed overlapping but also distinct sets of significant radiomics features across lung, mediastinal lymph node, and bone metastases. These results highlight the complementary biological information provided by FAPI and FDG. FAPI emphasizes stromal fibroblast activation, while FDG reflects glycolytic tumor metabolism.

The distribution of cases in the LMM and per-lesion analyses showed notable differences. For example, Lung Only metastases included 25 and 3 cases, respectively, Mediastinal Only had 3 and 9, and Bone Only included 17 and 24 cases. Multi-site combinations and all three sites together further reveal the diverse metastatic behavior of NSCLC.

In lung lesions, features such as original_GLSZM_LowGrayLevelZoneEmphasis and original_GLCM_ClusterProminence (LMM), and original_GLSZM_GrayLevelNonUniformity (per-lesion), suggest hypoxic and heterogeneous tumor regions. These patterns reflect aggressive growth with elevated stromal activity in FAPI and high metabolic demand in FDG.

Mediastinal-only metastases were marked by original_GLDM_LowGrayLevelEmphasis and original_FirstOrder_Kurtosis (LMM), as well as original_GLRLM_LongRunLowGrayLevelEmphasis (per-lesion). This indicates dense, high-contrast nodal regions, likely reflecting lymphoid infiltration or vascular remodeling.

Bone-only metastases showed features like original_GLDM_DependenceEntropy and original_GLDM_Correlation (LMM), and original_GLDM_GrayLevelNonUniformity (per-lesion). These point to complex tumor-bone interactions, including osteolytic or osteoblastic processes. FAPI captures stromal remodeling, while FDG highlights metabolic hotspots.

Multi-site metastatic groups further demonstrate the synergy of dual-tracer imaging. Lung and mediastinal lesions, with features such as original_FirstOrder_Mean (LMM) and original_GLDM_DependenceVariance (per-lesion), suggest shared inflammatory or vascular pathways. Lung and bone lesions, indicated by original_FirstOrder_10Percentile (LMM) and original_GLRLM_GrayLevelNonUniformityNormalized (per-lesion), reflect adaptive tumor phenotypes across pulmonary and osseous niches. Mediastinal and bone lesions, prominent in the per-lesion analysis, show high-energy patterns with original_GLDM_LargeDependenceHighGrayLevelEmphasis (LMM) and original_GLCM_SumSquares (per-lesion). The all-three-sites group, with original_GLRLM_LongRunEmphasis (LMM) and original_FirstOrder_RootMeanSquared (per-lesion), indicates aggressive and uniform tumor patterns, highlighting concordant stromal and metabolic signatures.

The paired per-lesion analysis complements the LMM model by providing high-resolution, site-specific insights into tumor-stroma-metabolism dynamics. This dual-tracer radiomics framework non-invasively captures the biological complexity of NSCLC metastases. It also offers a robust tool for prognostic stratification and may guide precision therapies targeting stromal remodeling or metabolic pathways. This approach could potentially improve outcomes for NSCLC patients with diverse metastatic patterns.

**Comparison with Existing Literature**

Prior studies have compared FAPI and FDG in NSCLC primarily via SUV metrics, reporting superior FAPI sensitivity in low-FDG-avid lesions (e.g., well-differentiated subtypes) and reduced background in inflammation(23-28). Our radiomics approach extends this by quantifying heterogeneity, aligning with emerging evidence that texture features predict treatment response; for instance, GLCM contrast in FDG PET correlates with EGFR mutation status and prognosis in NSCLC (41). Limited FAPI radiomics data exist, but in other cancers (e.g., gastric), FAPI texture features reflect stromal density and correlate with fibrosis scores on histopathology (42). Our findings of higher FAPI variance in lung lesions echo these, suggesting biological ties to CAF abundance, which histologically associates with poor outcomes (43). The per-lesion focus addresses gaps in patient-level analyses, revealing tissue-specific patterns absent in aggregate studies.

**Clinical Implications and Potential Applications**

The complementary radiomics profiles suggest hybrid FAPI-FDG approaches could enhance NSCLC management. For example, FAPI's stromal insights may improve delineation of fibrotic borders for radiotherapy planning, while FDG's metabolic focus aids in identifying aggressive clones (44). In personalized medicine, features like FAPI run entropy could predict response to CAF-targeted therapies (e.g., FAP inhibitors) or immunotherapies, as stromal heterogeneity modulates PD-L1 expression (45). Staging changes in 75% of cases (up/down) highlight FAPI's potential to refine TNM classification, particularly in bone and nodal sites where FDG specificity is limited by infection (46).

**Strengths, Limitations, and Future Directions**

Strengths of this study include rigorous per-lesion pairing of a total of 81 co-localized lesions (Lung: 21, Mediastinal lymph nodes: 42, Bone: 18). We applied IBSI-compliant preprocessing and used linear mixed models (LMM) to account for patient-level clustering, which helps ensure statistically robust and unbiased comparisons. To our knowledge, this is the first study to perform a direct, paired per-lesion radiomics comparison between [68Ga]-FAPI-46 and [18F]-FDG PET across multiple metastatic sites in NSCLC. It provides high-resolution insight into both stromal and metabolic heterogeneity.

Limitations include the relatively small number of patients (n = 14). However, the study analyzed a large number of lesions and performed rigorous per-lesion paired analysis, which improved statistical power despite the limited cohort. Being a pilot study, it focused on co-localized lesions, which may have excluded modality-specific sites. While semi-automatic segmentation with expert correction was applied to minimize variability, some observed differences in radiomics features may still be influenced by technical factors, including respiratory motion artifacts (RMA). Such artifacts can affect lesion delineation and intensity measurements, particularly in lung lesions, and should be carefully considered when interpreting modality-specific differences.. Future automation via AI could further standardize this process (47) .Biological validation via correlative histopathology (e.g., FAP staining) was not performed, representing a key next step.

Future studies should expand patient cohorts and include longitudinal data to monitor therapy response. Integrating multi-omics, such as genomics, could help link radiomics features to molecular drivers—for example, KRAS mutations that enhance glycolysis or TGF-β pathways that amplify stromal FAP (48). AI-driven automation for segmentation, along with machine-learning models combining dual-tracer radiomics, may further improve predictive power and accelerate translation into precision oncology.

## Conclusion

this pilot study demonstrates that FAPI and FDG radiomics provide biologically distinct windows into NSCLC heterogeneity, with FAPI emphasizing stromal dynamics and FDG metabolic activity. These findings support the development of integrated imaging strategies, which warrant validation in larger prospective studies.

**Ethics:** The study received approval from the Ethics Committee of SBMU (Registration No: IR.SBMU.NRITLD.REC.1401.073) and was registered in the Iranian Registry of Clinical Trials (https://irct.behdasht.gov.ir/trial/65361).

**Conflict of Interest:** Andrea Corsi is salaried employee and has stock options of the company radiomics.bio (Oncoradiomics SA). The rest of the Authors do not have conflicts to declare.

**Funding:** None

# Supplementary 01

| Tissue | Feature | LMM P-value | LMM Coef (Modality) | LMM Effect Size | N Observations | Adjusted LMM P-value |
|---|---|---|---|---|---|---|
| Lung | original_firstorder_10Percentile | 5.19E-34 | -3.502081535 | -12.15817338 | 42 | 4.83E-32 |
| Lung | original_firstorder_90Percentile | 1.68E-07 | -2.499105572 | -5.231451431 | 42 | 1.74E-06 |
| Lung | original_firstorder_Energy | 0.011557018 | -172.0620613 | -2.525389309 | 42 | 0.028284282 |
| Lung | original_firstorder_Entropy | 0.538438683 | -0.181987346 | -0.615175717 | 42 | 0.618207377 |
| Lung | original_firstorder_InterquartileRange | 0.001474461 | 6.115568998 | 3.179663664 | 42 | 0.00604904 |
| Lung | original_firstorder_Kurtosis | 0.163344018 | -0.085457065 | -1.393912538 | 42 | 0.223396965 |
| Lung | original_firstorder_Maximum | 0.002158944 | -2.156280271 | -3.067448885 | 42 | 0.007436362 |
| Lung | original_firstorder_MeanAbsoluteDeviation | 0.003277611 | 5.326411459 | 2.940426163 | 42 | 0.009832833 |
| Lung | original_firstorder_Mean | 1.86E-12 | -2.976839762 | -7.044738519 | 42 | 3.46E-11 |
| Lung | original_firstorder_Median | 1.57E-17 | -2.951963482 | -8.521575944 | 42 | 4.88E-16 |
| Lung | original_firstorder_Minimum | 4.88E-27 | -3.267469682 | -10.76796272 | 42 | 2.27E-25 |
| Lung | original_firstorder_Range | 0.040996169 | 3.069905717 | 2.043568752 | 42 | 0.071936674 |
| Lung | original_firstorder_RobustMeanAbsoluteDeviation | 0.001616023 | 6.015206472 | 3.15299914 | 42 | 0.006066226 |
| Lung | original_firstorder_RootMeanSquared | 2.94E-11 | -2.932627335 | -6.649780655 | 42 | 4.55E-10 |
| Lung | original_firstorder_Skewness | 0.278689713 | -1.156784281 | -1.083267486 | 42 | 0.35024518 |
| Lung | original_firstorder_TotalEnergy | 0.011557018 | -172.0620613 | -2.525389309 | 42 | 0.028284282 |
| Lung | original_firstorder_Uniformity | 0.003490269 | -0.067040493 | -2.920895439 | 42 | 0.010143596 |
| Lung | original_firstorder_Variance | 0.014064327 | 133.7324136 | 2.455616215 | 42 | 0.03190201 |
| Lung | original_glcm_Autocorrelation | 0.058769703 | 16.73361348 | 1.889912201 | 42 | 0.095887411 |
| Lung | original_glcm_ClusterProminence | 0.023750513 | 42334.34497 | 2.261141234 | 42 | 0.048017341 |
| Lung | original_glcm_ClusterShade | 0.912674273 | 825.2079026 | 0.109665991 | 42 | 0.953693341 |
| Lung | original_glcm_ClusterTendency | 0.018827153 | 92.21742988 | 2.34893579 | 42 | 0.039793756 |
| Lung | original_glcm_Contrast | 0.001630706 | 47.2704958 | 3.150357829 | 42 | 0.006066226 |
| Lung | original_glcm_Correlation | 0.068006493 | 0.105285093 | 1.824963793 | 42 | 0.107196676 |
| Lung | original_glcm_DifferenceAverage | 0.102108379 | 2.632606141 | 1.634717039 | 42 | 0.148376238 |
| Lung | original_glcm_DifferenceEntropy | 0.46669426 | 0.208084829 | 0.727868218 | 42 | 0.542532077 |
| Lung | original_glcm_DifferenceVariance | 0.00019042 | 24.82862256 | 3.731398271 | 42 | 0.001264932 |
| Lung | original_glcm_Id | 6.84E-06 | -0.108220658 | -4.498466312 | 42 | 5.79E-05 |
| Lung | original_glcm_Idm | 2.38E-06 | -0.133263095 | -4.718059738 | 42 | 2.21E-05 |
| Lung | original_glcm_Idmn | 0.925065333 | 0.000237875 | 0.094055168 | 42 | 0.955900844 |
| Lung | original_glcm_Idn | 0.195357767 | 0.00796053 | 1.294891194 | 42 | 0.255891159 |
| Lung | original_glcm_Imc1 | 0.018424949 | 0.093637251 | 2.356965687 | 42 | 0.039793756 |
| Lung | original_glcm_Imc2 | 0.948020998 | 0.002707266 | 0.065192167 | 42 | 0.958325574 |
| Lung | original_glcm_InverseVariance | 1.07E-13 | -1.345763843 | -7.432087964 | 42 | 2.49E-12 |
| Lung | original_glcm_JointAverage | 0.033776008 | 1.30341361 | 2.122735671 | 42 | 0.064105485 |
| Lung | original_glcm_JointEnergy | 0.008495656 | -0.02325043 | -2.631709142 | 42 | 0.021947111 |
| Lung | original_glcm_JointEntropy | 0.387547095 | -0.272322385 | -0.864074265 | 42 | 0.468076362 |
| Lung | original_glcm_MCC | 0.061698402 | 0.08574465 | 1.868456977 | 42 | 0.098930196 |
| Lung | original_glcm_MaximumProbability | 0.002613979 | -0.035921306 | -3.009825394 | 42 | 0.008218613 |
| Lung | original_glcm_SumAverage | 0.033776008 | 1.30341361 | 2.122735671 | 42 | 0.064105485 |
| Lung | original_glcm_SumEntropy | 0.214777042 | -0.283775761 | -1.240536447 | 42 | 0.277420346 |
| Lung | original_glcm_SumSquares | 0.013692148 | 83.69334117 | 2.465240058 | 42 | 0.031853459 |
| Lung | original_gldm_DependenceEntropy | 0.669401017 | -0.01909032 | -0.426970222 | 42 | 0.741122555 |
| Lung | original_gldm_DependenceNonUniformity | 0.005674871 | 9.905788087 | 2.765998988 | 42 | 0.015078943 |
| Lung | original_gldm_DependenceNonUniformityNormalized | 9.98E-09 | 1.538099323 | 5.731030659 | 42 | 1.33E-07 |
| Lung | original_gldm_DependenceVariance | 0.00194217 | -0.068605909 | -3.09893598 | 42 | 0.006946992 |
| Lung | original_gldm_GrayLevelNonUniformity | 0.286121619 | 0.162649876 | 1.066668361 | 42 | 0.354790808 |
| Lung | original_gldm_GrayLevelVariance | 0.02592867 | 92.28713688 | 2.227278451 | 42 | 0.051305666 |
| Lung | original_gldm_HighGrayLevelEmphasis | 0.043373703 | 16.02406209 | 2.020093372 | 42 | 0.073340989 |
| Lung | original_gldm_LargeDependenceEmphasis | 0.042177638 | -0.024598258 | -2.031763167 | 42 | 0.072639265 |
| Lung | original_gldm_LargeDependenceHighGrayLevelEmphasis | 0.941466617 | -0.033052835 | -0.073426643 | 42 | 0.958325574 |
| Lung | original_gldm_LargeDependenceLowGrayLevelEmphasis | 0.089316507 | -0.004411907 | -1.699014204 | 42 | 0.136171069 |
| Lung | original_gldm_LowGrayLevelEmphasis | 0.138707872 | -0.045827953 | -1.480619966 | 42 | 0.198458955 |
| Lung | original_gldm_SmallDependenceEmphasis | 0.023729229 | 10.28831977 | 2.261485179 | 42 | 0.048017341 |
| Lung | original_gldm_SmallDependenceHighGrayLevelEmphasis | 0.039150355 | 737.9761646 | 2.062603076 | 42 | 0.071391824 |
| Lung | original_gldm_SmallDependenceLowGrayLevelEmphasis | 0.092160746 | -2.294716138 | -1.684108271 | 42 | 0.138241119 |
| Lung | original_glrlm_GrayLevelNonUniformity | 0.982413507 | 0.005542505 | 0.022043185 | 42 | 0.982413507 |
| Lung | original_glrlm_GrayLevelNonUniformityNormalized | 0.000206323 | -0.087920428 | -3.711146356 | 42 | 0.001279201 |
| Lung | original_glrlm_GrayLevelVariance | 0.000850046 | 88.89456536 | 3.335962522 | 42 | 0.004391905 |
| Lung | original_glrlm_HighGrayLevelRunEmphasis | 0.040075778 | 15.85512965 | 2.052967002 | 42 | 0.071673988 |
| Lung | original_glrlm_LongRunEmphasis | 0.001169166 | 0.027853633 | 3.246297451 | 42 | 0.00572276 |
| Lung | original_glrlm_LongRunHighGrayLevelEmphasis | 0.004549941 | 2.722643409 | 2.837282001 | 42 | 0.012445426 |
| Lung | original_glrlm_LongRunLowGrayLevelEmphasis | 0.862056857 | 0.000513309 | 0.173756469 | 42 | 0.91103736 |
| Lung | original_glrlm_LowGrayLevelRunEmphasis | 0.077916215 | -0.060740409 | -1.762906769 | 42 | 0.120770134 |
| Lung | original_glrlm_RunEntropy | 4.77E-05 | 0.205111605 | 4.066634182 | 42 | 0.000341217 |
| Lung | original_glrlm_RunLengthNonUniformity | 0.849121161 | -2.751387908 | -0.190239882 | 42 | 0.91103736 |
| Lung | original_glrlm_RunLengthNonUniformityNormalized | 0.002651166 | -0.46856359 | -3.005532155 | 42 | 0.008218613 |
| Lung | original_glrlm_RunPercentage | 1.08E-05 | -0.350839627 | -4.400536435 | 42 | 8.37E-05 |
| Lung | original_glrlm_RunVariance | 0.860898395 | 0.001770644 | 0.175230658 | 42 | 0.91103736 |
| Lung | original_glrlm_ShortRunEmphasis | 0.000309271 | -0.231835712 | -3.607408332 | 42 | 0.001797636 |
| Lung | original_glrlm_ShortRunHighGrayLevelEmphasis | 0.149434355 | 29.17206042 | 1.441532315 | 42 | 0.210566591 |
| Lung | original_glrlm_ShortRunLowGrayLevelEmphasis | 0.015083829 | -0.243753309 | -2.43036007 | 42 | 0.033399908 |
| Lung | original_glszm_GrayLevelNonUniformity | 0.002348547 | 10.84890253 | 3.042201344 | 42 | 0.007800533 |
| Lung | original_glszm_GrayLevelNonUniformityNormalized | 0.00034122 | -0.130635162 | -3.581811319 | 42 | 0.001866675 |
| Lung | original_glszm_GrayLevelVariance | 0.013700412 | 30.08576811 | 2.465023873 | 42 | 0.031853459 |
| Lung | original_glszm_HighGrayLevelZoneEmphasis | 0.004348568 | 21.45932942 | 2.851703226 | 42 | 0.012255055 |
| Lung | original_glszm_LargeAreaEmphasis | 0.175718826 | 0.139337241 | 1.354055016 | 42 | 0.235618532 |
| Lung | original_glszm_LargeAreaHighGrayLevelEmphasis | 0.43115659 | 0.799924953 | 0.787214022 | 42 | 0.507564087 |
| Lung | original_glszm_LargeAreaLowGrayLevelEmphasis | 0.161631899 | 0.014522538 | 1.399604122 | 42 | 0.223396965 |
| Lung | original_glszm_LowGrayLevelZoneEmphasis | 0.001495999 | -0.172004209 | -3.17545853 | 42 | 0.00604904 |
| Lung | original_glszm_SizeZoneNonUniformity | 0.001433184 | 418.0503528 | 3.187883161 | 42 | 0.00604904 |
| Lung | original_glszm_SizeZoneNonUniformityNormalized | 1.34E-08 | 0.87133714 | 5.680247675 | 42 | 1.56E-07 |
| Lung | original_glszm_SmallAreaEmphasis | 0.046523426 | -70.0821729 | -1.990613218 | 42 | 0.077262118 |
| Lung | original_glszm_SmallAreaHighGrayLevelEmphasis | 0.614914004 | -441.2025095 | -0.503071499 | 42 | 0.689000029 |
| Lung | original_glszm_SmallAreaLowGrayLevelEmphasis | 0.100942807 | -12.56104724 | -1.640299991 | 42 | 0.148376238 |
| Lung | original_glszm_ZoneEntropy | 0.240330064 | 0.163200121 | 1.1741622 | 42 | 0.306173917 |
| Lung | original_glszm_ZonePercentage | 0.001352203 | 27.56835199 | 3.204663933 | 42 | 0.00604904 |
| Lung | original_glszm_ZoneVariance | 0.177347282 | 0.157276541 | 1.348967883 | 42 | 0.235618532 |
| Lung | original_ngtdm_Busyness | 0.552512094 | -0.00896957 | -0.594000029 | 42 | 0.62662957 |
| Lung | original_ngtdm_Coarseness | 0.37465657 | -0.445010706 | -0.887784708 | 42 | 0.458461329 |
| Lung | original_ngtdm_Complexity | 0.417564837 | 162.7163397 | 0.810653233 | 42 | 0.497865768 |
| Lung | original_ngtdm_Contrast | 0.036037069 | 5.406225043 | 2.096508923 | 42 | 0.067028949 |
| Lung | original_ngtdm_Strength | 0.802453079 | -6.932900367 | -0.250173627 | 42 | 0.877978075 |
| Mediastinal lymph node | original_firstorder_Energy | 0.002275483 | -16.30939223 | -3.051701483 | 84 | 0.005688707 |
| Mediastinal lymph node | original_firstorder_Kurtosis | 0.000250101 | 0.115363134 | 3.662156029 | 84 | 0.001042089 |
| Mediastinal lymph node | original_firstorder_TotalEnergy | 0.002275483 | -16.30939223 | -3.051701483 | 84 | 0.005688707 |
| Mediastinal lymph node | original_glcm_Autocorrelation | 0.011425122 | -10.36089033 | -2.529419978 | 84 | 0.022313209 |
| Mediastinal lymph node | original_glcm_Contrast | 0.006322688 | 18.30034445 | 2.730559695 | 84 | 0.014369746 |
| Mediastinal lymph node | original_glcm_DifferenceAverage | 0.733483901 | 0.283952745 | 0.340494849 | 84 | 0.833504433 |
| Mediastinal lymph node | original_glcm_DifferenceVariance | 0.000596482 | 10.22528222 | 3.43320885 | 84 | 0.002130294 |
| Mediastinal lymph node | original_glcm_JointEnergy | 0.285460811 | -0.008879072 | -1.06813236 | 84 | 0.339834299 |
| Mediastinal lymph node | original_gldm_GrayLevelNonUniformity | 0.96737108 | -0.000370632 | -0.040905692 | 84 | 0.96737108 |
| Mediastinal lymph node | original_gldm_HighGrayLevelEmphasis | 0.011602869 | -8.844197038 | -2.5239977 | 84 | 0.022313209 |
| Mediastinal lymph node | original_gldm_LargeDependenceHighGrayLevelEmphasis | 3.94E-08 | -0.787362307 | -5.493668428 | 84 | 3.28E-07 |
| Mediastinal lymph node | original_gldm_LowGrayLevelEmphasis | 0.000198016 | 0.106680698 | 3.721534509 | 84 | 0.000990079 |
| Mediastinal lymph node | original_glrlm_GrayLevelNonUniformityNormalized | 0.15372481 | -0.029471523 | -1.426497423 | 84 | 0.213506681 |
| Mediastinal lymph node | original_glrlm_HighGrayLevelRunEmphasis | 0.046883847 | -6.282188097 | -1.987347994 | 84 | 0.078139744 |
| Mediastinal lymph node | original_glrlm_LongRunEmphasis | 0.000714534 | 0.013802956 | 3.383940754 | 84 | 0.002232919 |
| Mediastinal lymph node | original_glrlm_LongRunHighGrayLevelEmphasis | 0.024705997 | -0.79141486 | -2.245968943 | 84 | 0.044117852 |
| Mediastinal lymph node | original_glrlm_RunLengthNonUniformity | 0.940107918 | -0.064043879 | -0.075134224 | 84 | 0.96737108 |
| Mediastinal lymph node | original_glrlm_RunLengthNonUniformityNormalized | 3.73E-07 | -0.484182412 | -5.08231828 | 84 | 2.33E-06 |
| Mediastinal lymph node | original_glrlm_RunPercentage | 5.78E-10 | -0.298131827 | -6.196196902 | 84 | 1.45E-08 |
| Mediastinal lymph node | original_glrlm_RunVariance | 0.067616695 | -0.005360174 | -1.827552799 | 84 | 0.105651086 |
| Mediastinal lymph node | original_glrlm_ShortRunEmphasis | 1.93E-08 | -0.214364411 | -5.618314618 | 84 | 2.41E-07 |
| Mediastinal lymph node | original_glrlm_ShortRunHighGrayLevelEmphasis | 0.074732182 | -13.35465568 | -1.782104565 | 84 | 0.109900267 |
| Mediastinal lymph node | original_glszm_GrayLevelNonUniformity | 0.852169225 | -0.117619618 | -0.186351371 | 84 | 0.926270896 |
| Mediastinal lymph node | original_glszm_HighGrayLevelZoneEmphasis | 0.211406005 | 3.757097007 | 1.249708627 | 84 | 0.264257506 |
| Mediastinal lymph node | original_ngtdm_Strength | 0.189933107 | -31.19551582 | -1.310777023 | 84 | 0.249911983 |
| Bone | original_firstorder_10Percentile | 3.40E-08 | -1.727584413 | -5.519529643 | 36 | 2.87E-07 |
| Bone | original_firstorder_90Percentile | 0.000385705 | -4.828763702 | -3.549676508 | 36 | 0.001630479 |
| Bone | original_firstorder_Energy | 0.011228118 | -12.68717697 | -2.535517931 | 36 | 0.028222026 |
| Bone | original_firstorder_Entropy | 6.73E-09 | -1.742553728 | -5.797528868 | 36 | 6.95E-08 |
| Bone | original_firstorder_InterquartileRange | 0.003436598 | -10.38771928 | -2.925720299 | 36 | 0.01183717 |
| Bone | original_firstorder_Kurtosis | 0.273753926 | 0.076728436 | 1.09445853 | 36 | 0.367073161 |
| Bone | original_firstorder_Maximum | 0.004664496 | -5.295312648 | -2.829334216 | 36 | 0.013993488 |
| Bone | original_firstorder_MeanAbsoluteDeviation | 0.035522361 | -11.31147337 | -2.102353041 | 36 | 0.07809144 |
| Bone | original_firstorder_Mean | 1.51E-09 | -3.173346669 | -6.043391773 | 36 | 3.11E-08 |
| Bone | original_firstorder_Median | 5.56E-14 | -2.783402487 | -7.51796879 | 36 | 5.17E-12 |
| Bone | original_firstorder_Minimum | 1.21E-06 | -1.33438593 | -4.854000711 | 36 | 8.66E-06 |
| Bone | original_firstorder_Range | 0.055001486 | -7.775240569 | -1.918864486 | 36 | 0.113669738 |
| Bone | original_firstorder_RobustMeanAbsoluteDeviation | 0.009348711 | -10.93106031 | -2.599031801 | 36 | 0.025571475 |
| Bone | original_firstorder_RootMeanSquared | 1.65E-06 | -3.683586689 | -4.791866021 | 36 | 1.05E-05 |
| Bone | original_firstorder_Skewness | 0.000792848 | -4.584668436 | -3.355279741 | 36 | 0.003011079 |
| Bone | original_firstorder_TotalEnergy | 0.011228118 | -12.68717697 | -2.535517931 | 36 | 0.028222026 |
| Bone | original_firstorder_Uniformity | 0.061053562 | 0.063086026 | 1.873107352 | 36 | 0.123434376 |
| Bone | original_firstorder_Variance | 0.227431822 | -845.6130582 | -1.207000835 | 36 | 0.349738199 |
| Bone | original_glcm_Autocorrelation | 0.099974544 | -29.38620547 | -1.644977047 | 36 | 0.19370068 |
| Bone | original_glcm_ClusterProminence | 0.310458841 | -3130070.386 | -1.014259715 | 36 | 0.389306376 |
| Bone | original_glcm_ClusterShade | 0.344893954 | 33815.49494 | 0.944539641 | 36 | 0.422041286 |
| Bone | original_glcm_ClusterTendency | 0.244601801 | -417.8928079 | -1.163561403 | 36 | 0.355436992 |
| Bone | original_glcm_Contrast | 0.134481793 | -59.406063 | -1.496659833 | 36 | 0.245231505 |
| Bone | original_glcm_Correlation | 1.69E-06 | -0.253680767 | -4.787748999 | 36 | 1.05E-05 |
| Bone | original_glcm_DifferenceAverage | 0.004026355 | -7.412646438 | -2.876089493 | 36 | 0.012912105 |
| Bone | original_glcm_DifferenceEntropy | 0.00080943 | -0.989654899 | -3.349549887 | 36 | 0.003011079 |
| Bone | original_glcm_DifferenceVariance | 0.233158799 | -24.12514173 | -1.192261312 | 36 | 0.349738199 |
| Bone | original_glcm_Id | 0.852231723 | 0.00582338 | 0.18627167 | 36 | 0.880639447 |
| Bone | original_glcm_Idm | 0.931633864 | -0.003189152 | -0.08578936 | 36 | 0.94176032 |
| Bone | original_glcm_Idmn | 0.011114823 | -0.006360664 | -2.53906795 | 36 | 0.028222026 |
| Bone | original_glcm_Idn | 0.561782155 | -0.00338753 | -0.580196439 | 36 | 0.637143175 |
| Bone | original_glcm_Imc1 | 0.014992734 | -0.231235203 | -2.432554517 | 36 | 0.036692744 |
| Bone | original_glcm_Imc2 | 4.69E-09 | -0.288706383 | -5.857762421 | 36 | 5.54E-08 |
| Bone | original_glcm_InverseVariance | 0.000110451 | -0.814759499 | -3.866406281 | 36 | 0.000540628 |
| Bone | original_glcm_JointAverage | 0.016868524 | -1.43712318 | -2.389560962 | 36 | 0.039219319 |
| Bone | original_glcm_JointEnergy | 0.848893591 | 0.006000084 | 0.190530314 | 36 | 0.880639447 |
| Bone | original_glcm_JointEntropy | 1.67E-09 | -1.732298356 | -6.027061284 | 36 | 3.11E-08 |
| Bone | original_glcm_MCC | 9.18E-06 | -0.222728312 | -4.435728083 | 36 | 5.02E-05 |
| Bone | original_glcm_MaximumProbability | 0.847543266 | 0.006679871 | 0.192253978 | 36 | 0.880639447 |
| Bone | original_glcm_SumAverage | 0.016868524 | -1.43712318 | -2.389560962 | 36 | 0.039219319 |
| Bone | original_glcm_SumEntropy | 3.71E-11 | -1.463159243 | -6.615275893 | 36 | 1.15E-09 |
| Bone | original_glcm_SumSquares | 0.232531536 | -341.6535029 | -1.193863068 | 36 | 0.349738199 |
| Bone | original_gldm_DependenceEntropy | 6.36E-13 | -0.236462916 | -7.192539027 | 36 | 2.96E-11 |
| Bone | original_gldm_DependenceNonUniformity | 0.004462494 | 0.438791596 | 2.843471959 | 36 | 0.01383373 |
| Bone | original_gldm_DependenceNonUniformityNormalized | 4.76E-09 | 1.362057855 | 5.85524548 | 36 | 5.54E-08 |
| Bone | original_gldm_DependenceVariance | 0.200270248 | -0.033669318 | -1.280782001 | 36 | 0.315680221 |
| Bone | original_gldm_GrayLevelNonUniformity | 0.005515804 | 0.020858371 | 2.775257372 | 36 | 0.015795236 |
| Bone | original_gldm_GrayLevelVariance | 0.244422421 | -381.9017335 | -1.164003926 | 36 | 0.355436992 |
| Bone | original_gldm_HighGrayLevelEmphasis | 0.143730756 | -32.2376894 | -1.46203815 | 36 | 0.256860277 |
| Bone | original_gldm_LargeDependenceEmphasis | 0.452292333 | -0.011111413 | -0.751598854 | 36 | 0.525789837 |
| Bone | original_gldm_LargeDependenceHighGrayLevelEmphasis | 0.005604761 | -0.455001613 | -2.770050454 | 36 | 0.015795236 |
| Bone | original_gldm_LargeDependenceLowGrayLevelEmphasis | 0.154668554 | -0.031536981 | -1.423233236 | 36 | 0.256860277 |
| Bone | original_gldm_LowGrayLevelEmphasis | 0.741468814 | -0.026918979 | -0.329908879 | 36 | 0.825606063 |
| Bone | original_gldm_SmallDependenceEmphasis | 0.284185673 | -4.75163707 | -1.070963875 | 36 | 0.367073161 |
| Bone | original_gldm_SmallDependenceHighGrayLevelEmphasis | 0.268459036 | -3476.074833 | -1.106618199 | 36 | 0.367073161 |
| Bone | original_gldm_SmallDependenceLowGrayLevelEmphasis | 0.003623977 | 3.427765491 | 2.909163155 | 36 | 0.01203678 |
| Bone | original_glrlm_GrayLevelNonUniformity | 0.44698483 | 0.015439284 | 0.760451365 | 36 | 0.525789837 |
| Bone | original_glrlm_GrayLevelNonUniformityNormalized | 0.090410116 | 0.059690471 | 1.693238304 | 36 | 0.178896613 |
| Bone | original_glrlm_GrayLevelVariance | 0.254194174 | -243.2974603 | -1.140221161 | 36 | 0.35972909 |
| Bone | original_glrlm_HighGrayLevelRunEmphasis | 0.154042417 | -38.27151837 | -1.425397199 | 36 | 0.256860277 |
| Bone | original_glrlm_LongRunEmphasis | 3.36E-07 | 0.04150898 | 5.101890271 | 36 | 2.61E-06 |
| Bone | original_glrlm_LongRunHighGrayLevelEmphasis | 0.313956755 | -1.033816752 | -1.00695426 | 36 | 0.389306376 |
| Bone | original_glrlm_LongRunLowGrayLevelEmphasis | 0.989881376 | -0.000181126 | -0.012682155 | 36 | 0.989881376 |
| Bone | original_glrlm_LowGrayLevelRunEmphasis | 0.831267761 | -0.017893209 | -0.213075902 | 36 | 0.880639447 |
| Bone | original_glrlm_RunEntropy | 0.019946553 | -0.092485212 | -2.327351719 | 36 | 0.045244621 |
| Bone | original_glrlm_RunLengthNonUniformity | 0.001594351 | -1.581261071 | -3.156938252 | 36 | 0.005702871 |
| Bone | original_glrlm_RunLengthNonUniformityNormalized | 2.61E-06 | -0.778030027 | -4.699244513 | 36 | 1.52E-05 |
| Bone | original_glrlm_RunPercentage | 3.63E-09 | -0.564109453 | -5.900189751 | 36 | 5.54E-08 |
| Bone | original_glrlm_RunVariance | 0.277813622 | 0.007804266 | 1.085243953 | 36 | 0.367073161 |
| Bone | original_glrlm_ShortRunEmphasis | 1.47E-05 | -0.29830677 | -4.333105921 | 36 | 7.60E-05 |
| Bone | original_glrlm_ShortRunHighGrayLevelEmphasis | 0.147054108 | -115.5060629 | -1.450015819 | 36 | 0.256860277 |
| Bone | original_glrlm_ShortRunLowGrayLevelEmphasis | 0.745708702 | -0.047846649 | -0.324302929 | 36 | 0.825606063 |
| Bone | original_glszm_GrayLevelNonUniformity | 0.18440089 | -0.531497649 | -1.327325931 | 36 | 0.295677288 |
| Bone | original_glszm_GrayLevelNonUniformityNormalized | 0.109576963 | 0.072298032 | 1.600097468 | 36 | 0.207972602 |
| Bone | original_glszm_GrayLevelVariance | 0.255291612 | -108.3351295 | -1.137590268 | 36 | 0.35972909 |
| Bone | original_glszm_HighGrayLevelZoneEmphasis | 0.265829098 | -37.34491233 | -1.112719081 | 36 | 0.367073161 |
| Bone | original_glszm_LargeAreaEmphasis | 0.16205881 | 0.001606844 | 1.398180704 | 36 | 0.264411742 |
| Bone | original_glszm_LargeAreaHighGrayLevelEmphasis | 0.771493706 | 0.001518644 | 0.29042162 | 36 | 0.844104878 |
| Bone | original_glszm_LargeAreaLowGrayLevelEmphasis | 0.41680092 | 0.000480877 | 0.811983813 | 36 | 0.496954943 |
| Bone | original_glszm_LowGrayLevelZoneEmphasis | 0.123383007 | 0.098449049 | 1.540727967 | 36 | 0.229492393 |
| Bone | original_glszm_SizeZoneNonUniformity | 0.844167095 | 1.196034468 | 0.196566108 | 36 | 0.880639447 |
| Bone | original_glszm_SizeZoneNonUniformityNormalized | 0.000344017 | 0.651769028 | 3.579678908 | 36 | 0.001523502 |
| Bone | original_glszm_SmallAreaEmphasis | 2.85E-08 | 33.34479913 | 5.5502876 | 36 | 2.65E-07 |
| Bone | original_glszm_SmallAreaHighGrayLevelEmphasis | 0.549778669 | 438.0884461 | 0.598091818 | 36 | 0.631227361 |
| Bone | original_glszm_SmallAreaLowGrayLevelEmphasis | 0.000797372 | 5.697133438 | 3.35370554 | 36 | 0.003011079 |
| Bone | original_glszm_ZoneEntropy | 0.000290337 | -0.367120033 | -3.623772272 | 36 | 0.001350068 |
| Bone | original_glszm_ZonePercentage | 0.046839529 | -15.65770553 | -1.987748348 | 36 | 0.099001731 |
| Bone | original_glszm_ZoneVariance | 0.15234106 | 0.001502266 | 1.43131115 | 36 | 0.256860277 |
| Bone | original_ngtdm_Busyness | 0.036106795 | 0.015506602 | 2.095722725 | 36 | 0.07809144 |
| Bone | original_ngtdm_Coarseness | 0.862997708 | -0.184942823 | -0.172559475 | 36 | 0.88196469 |
| Bone | original_ngtdm_Complexity | 0.291115715 | -1550.273883 | -1.055677206 | 36 | 0.370873445 |
| Bone | original_ngtdm_Contrast | 0.410699633 | -3.758734068 | -0.82266305 | 36 | 0.496039817 |
| Bone | original_ngtdm_Strength | 0.28250776 | -1235.091808 | -1.074702911 | 36 | 0.367073161 |

# Supplementary 02

| | Lung Only | Mediastinal Only | Bone Only | Lung and Mediastinal | Lung and Bone | Mediastinal and Bone | All Three |
|---|---|---|---|---|---|---|---|
| **Count** | 3 | 9 | 24 | 6 | 2 | 25 | 15 |
| **Features** | original_glszm_GrayLevelNonUniformity, original_glcm_MaximumProbability, original_glszm_SizeZoneNonUniformity | original_glrlm_LongRunLowGrayLevelEmphasis, original_ngtdm_Busyness, original_firstorder_Kurtosis, original_glrlm_ShortRunLowGrayLevelEmphasis, original_glszm_SmallAreaHighGrayLevelEmphasis, original_gldm_LowGrayLevelEmphasis, original_glrlm_LowGrayLevelRunEmphasis, original_ngtdm_Contrast, original_gldm_LargeDependenceLowGrayLevelEmphasis | original_gldm_GrayLevelNonUniformity, original_glcm_Correlation, original_glrlm_RunLengthNonUniformity, original_glcm_DifferenceAverage, original_ngtdm_Strength, original_glszm_ZoneEntropy, original_glcm_Imc2, original_firstorder_Entropy, original_firstorder_InterquartileRange, original_glcm_JointEnergy, original_ngtdm_Complexity, original_firstorder_Skewness, original_firstorder_Variance, original_firstorder_Range, original_glszm_HighGrayLevelZoneEmphasis, original_glszm_ZonePercentage, original_firstorder_MeanAbsoluteDeviation, original_glcm_JointEntropy, original_glszm_GrayLevelVariance, original_glcm_MCC, original_glcm_DifferenceEntropy, original_glszm_GrayLevelNon | original_gldm_DependenceVariance, original_glcm_DifferenceVariance, original_gldm_LargeDependenceEmphasis, original_glszm_LargeAreaHighGrayLevelEmphasis, original_glcm_Idm, original_glcm_Id | original_glrlm_GrayLevelNonUniformityNormalized, original_firstorder_Uniformity | original_glcm_SumSquares, original_glcm_SumAverage, original_glcm_Contrast, original_firstorder_TotalEnergy, original_glrlm_HighGrayLevelRunEmphasis, original_gldm_DependenceEntropy, original_glcm_ClusterProminence, original_glrlm_LongRunHighGrayLevelEmphasis, original_gldm_SmallDependenceLowGrayLevelEmphasis, original_glcm_Autocorrelation, original_gldm_LargeDependenceHighGrayLevelEmphasis, original_glcm_JointAverage, original_glrlm_RunLengthNonUniformityNormalized, original_glcm_SumEntropy, original_firstorder_Maximum, original_glrlm_ShortRunHighGrayLevelEmphasis, original_glszm_SmallAreaEmphasis, original_gldm_GrayLevelVaria | original_firstorder_Mean, original_firstorder_RootMeanSquared, original_gldm_DependenceNonUniformity, original_glrlm_RunEntropy, original_glszm_SmallAreaLowGrayLevelEmphasis, original_firstorder_10Percentile, original_glrlm_LongRunEmphasis, original_glrlm_GrayLevelVariance, original_gldm_DependenceNonUniformityNormalized, original_glszm_SizeZoneNonUniformityNormalized, original_firstorder_Minimum, original_glrlm_RunPercentage, original_firstorder_Median, original_glcm_InverseVariance, original_firstorder_90Percentile |

# Supplementary 03

| Tissue | Feature | LMM P-value | LMM Coef (Modality) | LMM Effect Size | N Observations | Adjusted LMM P-value |
|---|---|---|---|---|---|---|
| Lung | original_firstorder_10Percentile | 5.19E-34 | -3.502081535 | -12.15817338 | 42 | 4.83E-32 |
| Lung | original_firstorder_90Percentile | 1.68E-07 | -2.499105572 | -5.231451431 | 42 | 1.74E-06 |
| Lung | original_firstorder_Energy | 0.011557018 | -172.0620613 | -2.525389309 | 42 | 0.028284282 |
| Lung | original_firstorder_Entropy | 0.538438683 | -0.181987346 | -0.615175717 | 42 | 0.618207377 |
| Lung | original_firstorder_InterquartileRange | 0.001474461 | 6.115568998 | 3.179663664 | 42 | 0.00604904 |
| Lung | original_firstorder_Kurtosis | 0.163344018 | -0.085457065 | -1.393912538 | 42 | 0.223396965 |
| Lung | original_firstorder_Maximum | 0.002158944 | -2.156280271 | -3.067448885 | 42 | 0.007436362 |
| Lung | original_firstorder_MeanAbsoluteDeviation | 0.003277611 | 5.326411459 | 2.940426163 | 42 | 0.009832833 |
| Lung | original_firstorder_Mean | 1.86E-12 | -2.976839762 | -7.044738519 | 42 | 3.46E-11 |
| Lung | original_firstorder_Median | 1.57E-17 | -2.951963482 | -8.521575944 | 42 | 4.88E-16 |
| Lung | original_firstorder_Minimum | 4.88E-27 | -3.267469682 | -10.76796272 | 42 | 2.27E-25 |
| Lung | original_firstorder_Range | 0.040996169 | 3.069905717 | 2.043568752 | 42 | 0.071936674 |
| Lung | original_firstorder_RobustMeanAbsoluteDeviation | 0.001616023 | 6.015206472 | 3.15299914 | 42 | 0.006066226 |
| Lung | original_firstorder_RootMeanSquared | 2.94E-11 | -2.932627335 | -6.649780655 | 42 | 4.55E-10 |
| Lung | original_firstorder_Skewness | 0.278689713 | -1.156784281 | -1.083267486 | 42 | 0.35024518 |
| Lung | original_firstorder_TotalEnergy | 0.011557018 | -172.0620613 | -2.525389309 | 42 | 0.028284282 |
| Lung | original_firstorder_Uniformity | 0.003490269 | -0.067040493 | -2.920895439 | 42 | 0.010143596 |
| Lung | original_firstorder_Variance | 0.014064327 | 133.7324136 | 2.455616215 | 42 | 0.03190201 |
| Lung | original_glcm_Autocorrelation | 0.058769703 | 16.73361348 | 1.889912201 | 42 | 0.095887411 |
| Lung | original_glcm_ClusterProminence | 0.023750513 | 42334.34497 | 2.261141234 | 42 | 0.048017341 |
| Lung | original_glcm_ClusterShade | 0.912674273 | 825.2079026 | 0.109665991 | 42 | 0.953693341 |
| Lung | original_glcm_ClusterTendency | 0.018827153 | 92.21742988 | 2.34893579 | 42 | 0.039793756 |
| Lung | original_glcm_Contrast | 0.001630706 | 47.2704958 | 3.150357829 | 42 | 0.006066226 |
| Lung | original_glcm_Correlation | 0.068006493 | 0.105285093 | 1.824963793 | 42 | 0.107196676 |
| Lung | original_glcm_DifferenceAverage | 0.102108379 | 2.632606141 | 1.634717039 | 42 | 0.148376238 |
| Lung | original_glcm_DifferenceEntropy | 0.46669426 | 0.208084829 | 0.727868218 | 42 | 0.542532077 |
| Lung | original_glcm_DifferenceVariance | 0.00019042 | 24.82862256 | 3.731398271 | 42 | 0.001264932 |
| Lung | original_glcm_Id | 6.84E-06 | -0.108220658 | -4.498466312 | 42 | 5.79E-05 |
| Lung | original_glcm_Idm | 2.38E-06 | -0.133263095 | -4.718059738 | 42 | 2.21E-05 |
| Lung | original_glcm_Idmn | 0.925065333 | 0.000237875 | 0.094055168 | 42 | 0.955900844 |
| Lung | original_glcm_Idn | 0.195357767 | 0.00796053 | 1.294891194 | 42 | 0.255891159 |
| Lung | original_glcm_Imc1 | 0.018424949 | 0.093637251 | 2.356965687 | 42 | 0.039793756 |
| Lung | original_glcm_Imc2 | 0.948020998 | 0.002707266 | 0.065192167 | 42 | 0.958325574 |
| Lung | original_glcm_InverseVariance | 1.07E-13 | -1.345763843 | -7.432087964 | 42 | 2.49E-12 |
| Lung | original_glcm_JointAverage | 0.033776008 | 1.30341361 | 2.122735671 | 42 | 0.064105485 |
| Lung | original_glcm_JointEnergy | 0.008495656 | -0.02325043 | -2.631709142 | 42 | 0.021947111 |
| Lung | original_glcm_JointEntropy | 0.387547095 | -0.272322385 | -0.864074265 | 42 | 0.468076362 |
| Lung | original_glcm_MCC | 0.061698402 | 0.08574465 | 1.868456977 | 42 | 0.098930196 |
| Lung | original_glcm_MaximumProbability | 0.002613979 | -0.035921306 | -3.009825394 | 42 | 0.008218613 |
| Lung | original_glcm_SumAverage | 0.033776008 | 1.30341361 | 2.122735671 | 42 | 0.064105485 |
| Lung | original_glcm_SumEntropy | 0.214777042 | -0.283775761 | -1.240536447 | 42 | 0.277420346 |
| Lung | original_glcm_SumSquares | 0.013692148 | 83.69334117 | 2.465240058 | 42 | 0.031853459 |
| Lung | original_gldm_DependenceEntropy | 0.669401017 | -0.01909032 | -0.426970222 | 42 | 0.741122555 |
| Lung | original_gldm_DependenceNonUniformity | 0.005674871 | 9.905788087 | 2.765998988 | 42 | 0.015078943 |
| Lung | original_gldm_DependenceNonUniformityNormalized | 9.98E-09 | 1.538099323 | 5.731030659 | 42 | 1.33E-07 |
| Lung | original_gldm_DependenceVariance | 0.00194217 | -0.068605909 | -3.09893598 | 42 | 0.006946992 |
| Lung | original_gldm_GrayLevelNonUniformity | 0.286121619 | 0.162649876 | 1.066668361 | 42 | 0.354790808 |
| Lung | original_gldm_GrayLevelVariance | 0.02592867 | 92.28713688 | 2.227278451 | 42 | 0.051305666 |
| Lung | original_gldm_HighGrayLevelEmphasis | 0.043373703 | 16.02406209 | 2.020093372 | 42 | 0.073340989 |
| Lung | original_gldm_LargeDependenceEmphasis | 0.042177638 | -0.024598258 | -2.031763167 | 42 | 0.072639265 |
| Lung | original_gldm_LargeDependenceHighGrayLevelEmphasis | 0.941466617 | -0.033052835 | -0.073426643 | 42 | 0.958325574 |
| Lung | original_gldm_LargeDependenceLowGrayLevelEmphasis | 0.089316507 | -0.004411907 | -1.699014204 | 42 | 0.136171069 |
| Lung | original_gldm_LowGrayLevelEmphasis | 0.138707872 | -0.045827953 | -1.480619966 | 42 | 0.198458955 |
| Lung | original_gldm_SmallDependenceEmphasis | 0.023729229 | 10.28831977 | 2.261485179 | 42 | 0.048017341 |
| Lung | original_gldm_SmallDependenceHighGrayLevelEmphasis | 0.039150355 | 737.9761646 | 2.062603076 | 42 | 0.071391824 |
| Lung | original_gldm_SmallDependenceLowGrayLevelEmphasis | 0.092160746 | -2.294716138 | -1.684108271 | 42 | 0.138241119 |
| Lung | original_glrlm_GrayLevelNonUniformity | 0.982413507 | 0.005542505 | 0.022043185 | 42 | 0.982413507 |
| Lung | original_glrlm_GrayLevelNonUniformityNormalized | 0.000206323 | -0.087920428 | -3.711146356 | 42 | 0.001279201 |
| Lung | original_glrlm_GrayLevelVariance | 0.000850046 | 88.89456536 | 3.335962522 | 42 | 0.004391905 |
| Lung | original_glrlm_HighGrayLevelRunEmphasis | 0.040075778 | 15.85512965 | 2.052967002 | 42 | 0.071673988 |
| Lung | original_glrlm_LongRunEmphasis | 0.001169166 | 0.027853633 | 3.246297451 | 42 | 0.00572276 |
| Lung | original_glrlm_LongRunHighGrayLevelEmphasis | 0.004549941 | 2.722643409 | 2.837282001 | 42 | 0.012445426 |
| Lung | original_glrlm_LongRunLowGrayLevelEmphasis | 0.862056857 | 0.000513309 | 0.173756469 | 42 | 0.91103736 |
| Lung | original_glrlm_LowGrayLevelRunEmphasis | 0.077916215 | -0.060740409 | -1.762906769 | 42 | 0.120770134 |
| Lung | original_glrlm_RunEntropy | 4.77E-05 | 0.205111605 | 4.066634182 | 42 | 0.000341217 |
| Lung | original_glrlm_RunLengthNonUniformity | 0.849121161 | -2.751387908 | -0.190239882 | 42 | 0.91103736 |
| Lung | original_glrlm_RunLengthNonUniformityNormalized | 0.002651166 | -0.46856359 | -3.005532155 | 42 | 0.008218613 |
| Lung | original_glrlm_RunPercentage | 1.08E-05 | -0.350839627 | -4.400536435 | 42 | 8.37E-05 |
| Lung | original_glrlm_RunVariance | 0.860898395 | 0.001770644 | 0.175230658 | 42 | 0.91103736 |
| Lung | original_glrlm_ShortRunEmphasis | 0.000309271 | -0.231835712 | -3.607408332 | 42 | 0.001797636 |
| Lung | original_glrlm_ShortRunHighGrayLevelEmphasis | 0.149434355 | 29.17206042 | 1.441532315 | 42 | 0.210566591 |
| Lung | original_glrlm_ShortRunLowGrayLevelEmphasis | 0.015083829 | -0.243753309 | -2.43036007 | 42 | 0.033399908 |
| Lung | original_glszm_GrayLevelNonUniformity | 0.002348547 | 10.84890253 | 3.042201344 | 42 | 0.007800533 |
| Lung | original_glszm_GrayLevelNonUniformityNormalized | 0.00034122 | -0.130635162 | -3.581811319 | 42 | 0.001866675 |
| Lung | original_glszm_GrayLevelVariance | 0.013700412 | 30.08576811 | 2.465023873 | 42 | 0.031853459 |
| Lung | original_glszm_HighGrayLevelZoneEmphasis | 0.004348568 | 21.45932942 | 2.851703226 | 42 | 0.012255055 |
| Lung | original_glszm_LargeAreaEmphasis | 0.175718826 | 0.139337241 | 1.354055016 | 42 | 0.235618532 |
| Lung | original_glszm_LargeAreaHighGrayLevelEmphasis | 0.43115659 | 0.799924953 | 0.787214022 | 42 | 0.507564087 |
| Lung | original_glszm_LargeAreaLowGrayLevelEmphasis | 0.161631899 | 0.014522538 | 1.399604122 | 42 | 0.223396965 |
| Lung | original_glszm_LowGrayLevelZoneEmphasis | 0.001495999 | -0.172004209 | -3.17545853 | 42 | 0.00604904 |
| Lung | original_glszm_SizeZoneNonUniformity | 0.001433184 | 418.0503528 | 3.187883161 | 42 | 0.00604904 |
| Lung | original_glszm_SizeZoneNonUniformityNormalized | 1.34E-08 | 0.87133714 | 5.680247675 | 42 | 1.56E-07 |
| Lung | original_glszm_SmallAreaEmphasis | 0.046523426 | -70.0821729 | -1.990613218 | 42 | 0.077262118 |
| Lung | original_glszm_SmallAreaHighGrayLevelEmphasis | 0.614914004 | -441.2025095 | -0.503071499 | 42 | 0.689000029 |
| Lung | original_glszm_SmallAreaLowGrayLevelEmphasis | 0.100942807 | -12.56104724 | -1.640299991 | 42 | 0.148376238 |
| Lung | original_glszm_ZoneEntropy | 0.240330064 | 0.163200121 | 1.1741622 | 42 | 0.306173917 |
| Lung | original_glszm_ZonePercentage | 0.001352203 | 27.56835199 | 3.204663933 | 42 | 0.00604904 |
| Lung | original_glszm_ZoneVariance | 0.177347282 | 0.157276541 | 1.348967883 | 42 | 0.235618532 |
| Lung | original_ngtdm_Busyness | 0.552512094 | -0.00896957 | -0.594000029 | 42 | 0.62662957 |
| Lung | original_ngtdm_Coarseness | 0.37465657 | -0.445010706 | -0.887784708 | 42 | 0.458461329 |
| Lung | original_ngtdm_Complexity | 0.417564837 | 162.7163397 | 0.810653233 | 42 | 0.497865768 |
| Lung | original_ngtdm_Contrast | 0.036037069 | 5.406225043 | 2.096508923 | 42 | 0.067028949 |
| Lung | original_ngtdm_Strength | 0.802453079 | -6.932900367 | -0.250173627 | 42 | 0.877978075 |
| Mediastinal lymph node | original_firstorder_Energy | 0.002275483 | -16.30939223 | -3.051701483 | 84 | 0.005688707 |
| Mediastinal lymph node | original_firstorder_Kurtosis | 0.000250101 | 0.115363134 | 3.662156029 | 84 | 0.001042089 |
| Mediastinal lymph node | original_firstorder_TotalEnergy | 0.002275483 | -16.30939223 | -3.051701483 | 84 | 0.005688707 |
| Mediastinal lymph node | original_glcm_Autocorrelation | 0.011425122 | -10.36089033 | -2.529419978 | 84 | 0.022313209 |
| Mediastinal lymph node | original_glcm_Contrast | 0.006322688 | 18.30034445 | 2.730559695 | 84 | 0.014369746 |
| Mediastinal lymph node | original_glcm_DifferenceAverage | 0.733483901 | 0.283952745 | 0.340494849 | 84 | 0.833504433 |
| Mediastinal lymph node | original_glcm_DifferenceVariance | 0.000596482 | 10.22528222 | 3.43320885 | 84 | 0.002130294 |
| Mediastinal lymph node | original_glcm_JointEnergy | 0.285460811 | -0.008879072 | -1.06813236 | 84 | 0.339834299 |
| Mediastinal lymph node | original_gldm_GrayLevelNonUniformity | 0.96737108 | -0.000370632 | -0.040905692 | 84 | 0.96737108 |
| Mediastinal lymph node | original_gldm_HighGrayLevelEmphasis | 0.011602869 | -8.844197038 | -2.5239977 | 84 | 0.022313209 |
| Mediastinal lymph node | original_gldm_LargeDependenceHighGrayLevelEmphasis | 3.94E-08 | -0.787362307 | -5.493668428 | 84 | 3.28E-07 |
| Mediastinal lymph node | original_gldm_LowGrayLevelEmphasis | 0.000198016 | 0.106680698 | 3.721534509 | 84 | 0.000990079 |
| Mediastinal lymph node | original_glrlm_GrayLevelNonUniformityNormalized | 0.15372481 | -0.029471523 | -1.426497423 | 84 | 0.213506681 |
| Mediastinal lymph node | original_glrlm_HighGrayLevelRunEmphasis | 0.046883847 | -6.282188097 | -1.987347994 | 84 | 0.078139744 |
| Mediastinal lymph node | original_glrlm_LongRunEmphasis | 0.000714534 | 0.013802956 | 3.383940754 | 84 | 0.002232919 |
| Mediastinal lymph node | original_glrlm_LongRunHighGrayLevelEmphasis | 0.024705997 | -0.79141486 | -2.245968943 | 84 | 0.044117852 |
| Mediastinal lymph node | original_glrlm_RunLengthNonUniformity | 0.940107918 | -0.064043879 | -0.075134224 | 84 | 0.96737108 |
| Mediastinal lymph node | original_glrlm_RunLengthNonUniformityNormalized | 3.73E-07 | -0.484182412 | -5.08231828 | 84 | 2.33E-06 |
| Mediastinal lymph node | original_glrlm_RunPercentage | 5.78E-10 | -0.298131827 | -6.196196902 | 84 | 1.45E-08 |
| Mediastinal lymph node | original_glrlm_RunVariance | 0.067616695 | -0.005360174 | -1.827552799 | 84 | 0.105651086 |
| Mediastinal lymph node | original_glrlm_ShortRunEmphasis | 1.93E-08 | -0.214364411 | -5.618314618 | 84 | 2.41E-07 |
| Mediastinal lymph node | original_glrlm_ShortRunHighGrayLevelEmphasis | 0.074732182 | -13.35465568 | -1.782104565 | 84 | 0.109900267 |
| Mediastinal lymph node | original_glszm_GrayLevelNonUniformity | 0.852169225 | -0.117619618 | -0.186351371 | 84 | 0.926270896 |
| Mediastinal lymph node | original_glszm_HighGrayLevelZoneEmphasis | 0.211406005 | 3.757097007 | 1.249708627 | 84 | 0.264257506 |
| Mediastinal lymph node | original_ngtdm_Strength | 0.189933107 | -31.19551582 | -1.310777023 | 84 | 0.249911983 |
| Bone | original_firstorder_10Percentile | 3.40E-08 | -1.727584413 | -5.519529643 | 36 | 2.87E-07 |
| Bone | original_firstorder_90Percentile | 0.000385705 | -4.828763702 | -3.549676508 | 36 | 0.001630479 |
| Bone | original_firstorder_Energy | 0.011228118 | -12.68717697 | -2.535517931 | 36 | 0.028222026 |
| Bone | original_firstorder_Entropy | 6.73E-09 | -1.742553728 | -5.797528868 | 36 | 6.95E-08 |
| Bone | original_firstorder_InterquartileRange | 0.003436598 | -10.38771928 | -2.925720299 | 36 | 0.01183717 |
| Bone | original_firstorder_Kurtosis | 0.273753926 | 0.076728436 | 1.09445853 | 36 | 0.367073161 |
| Bone | original_firstorder_Maximum | 0.004664496 | -5.295312648 | -2.829334216 | 36 | 0.013993488 |
| Bone | original_firstorder_MeanAbsoluteDeviation | 0.035522361 | -11.31147337 | -2.102353041 | 36 | 0.07809144 |
| Bone | original_firstorder_Mean | 1.51E-09 | -3.173346669 | -6.043391773 | 36 | 3.11E-08 |
| Bone | original_firstorder_Median | 5.56E-14 | -2.783402487 | -7.51796879 | 36 | 5.17E-12 |
| Bone | original_firstorder_Minimum | 1.21E-06 | -1.33438593 | -4.854000711 | 36 | 8.66E-06 |
| Bone | original_firstorder_Range | 0.055001486 | -7.775240569 | -1.918864486 | 36 | 0.113669738 |
| Bone | original_firstorder_RobustMeanAbsoluteDeviation | 0.009348711 | -10.93106031 | -2.599031801 | 36 | 0.025571475 |
| Bone | original_firstorder_RootMeanSquared | 1.65E-06 | -3.683586689 | -4.791866021 | 36 | 1.05E-05 |
| Bone | original_firstorder_Skewness | 0.000792848 | -4.584668436 | -3.355279741 | 36 | 0.003011079 |
| Bone | original_firstorder_TotalEnergy | 0.011228118 | -12.68717697 | -2.535517931 | 36 | 0.028222026 |
| Bone | original_firstorder_Uniformity | 0.061053562 | 0.063086026 | 1.873107352 | 36 | 0.123434376 |
| Bone | original_firstorder_Variance | 0.227431822 | -845.6130582 | -1.207000835 | 36 | 0.349738199 |
| Bone | original_glcm_Autocorrelation | 0.099974544 | -29.38620547 | -1.644977047 | 36 | 0.19370068 |
| Bone | original_glcm_ClusterProminence | 0.310458841 | -3130070.386 | -1.014259715 | 36 | 0.389306376 |
| Bone | original_glcm_ClusterShade | 0.344893954 | 33815.49494 | 0.944539641 | 36 | 0.422041286 |
| Bone | original_glcm_ClusterTendency | 0.244601801 | -417.8928079 | -1.163561403 | 36 | 0.355436992 |
| Bone | original_glcm_Contrast | 0.134481793 | -59.406063 | -1.496659833 | 36 | 0.245231505 |
| Bone | original_glcm_Correlation | 1.69E-06 | -0.253680767 | -4.787748999 | 36 | 1.05E-05 |
| Bone | original_glcm_DifferenceAverage | 0.004026355 | -7.412646438 | -2.876089493 | 36 | 0.012912105 |
| Bone | original_glcm_DifferenceEntropy | 0.00080943 | -0.989654899 | -3.349549887 | 36 | 0.003011079 |
| Bone | original_glcm_DifferenceVariance | 0.233158799 | -24.12514173 | -1.192261312 | 36 | 0.349738199 |
| Bone | original_glcm_Id | 0.852231723 | 0.00582338 | 0.18627167 | 36 | 0.880639447 |
| Bone | original_glcm_Idm | 0.931633864 | -0.003189152 | -0.08578936 | 36 | 0.94176032 |
| Bone | original_glcm_Idmn | 0.011114823 | -0.006360664 | -2.53906795 | 36 | 0.028222026 |
| Bone | original_glcm_Idn | 0.561782155 | -0.00338753 | -0.580196439 | 36 | 0.637143175 |
| Bone | original_glcm_Imc1 | 0.014992734 | -0.231235203 | -2.432554517 | 36 | 0.036692744 |
| Bone | original_glcm_Imc2 | 4.69E-09 | -0.288706383 | -5.857762421 | 36 | 5.54E-08 |
| Bone | original_glcm_InverseVariance | 0.000110451 | -0.814759499 | -3.866406281 | 36 | 0.000540628 |
| Bone | original_glcm_JointAverage | 0.016868524 | -1.43712318 | -2.389560962 | 36 | 0.039219319 |
| Bone | original_glcm_JointEnergy | 0.848893591 | 0.006000084 | 0.190530314 | 36 | 0.880639447 |
| Bone | original_glcm_JointEntropy | 1.67E-09 | -1.732298356 | -6.027061284 | 36 | 3.11E-08 |
| Bone | original_glcm_MCC | 9.18E-06 | -0.222728312 | -4.435728083 | 36 | 5.02E-05 |
| Bone | original_glcm_MaximumProbability | 0.847543266 | 0.006679871 | 0.192253978 | 36 | 0.880639447 |
| Bone | original_glcm_SumAverage | 0.016868524 | -1.43712318 | -2.389560962 | 36 | 0.039219319 |
| Bone | original_glcm_SumEntropy | 3.71E-11 | -1.463159243 | -6.615275893 | 36 | 1.15E-09 |
| Bone | original_glcm_SumSquares | 0.232531536 | -341.6535029 | -1.193863068 | 36 | 0.349738199 |
| Bone | original_gldm_DependenceEntropy | 6.36E-13 | -0.236462916 | -7.192539027 | 36 | 2.96E-11 |
| Bone | original_gldm_DependenceNonUniformity | 0.004462494 | 0.438791596 | 2.843471959 | 36 | 0.01383373 |
| Bone | original_gldm_DependenceNonUniformityNormalized | 4.76E-09 | 1.362057855 | 5.85524548 | 36 | 5.54E-08 |
| Bone | original_gldm_DependenceVariance | 0.200270248 | -0.033669318 | -1.280782001 | 36 | 0.315680221 |
| Bone | original_gldm_GrayLevelNonUniformity | 0.005515804 | 0.020858371 | 2.775257372 | 36 | 0.015795236 |
| Bone | original_gldm_GrayLevelVariance | 0.244422421 | -381.9017335 | -1.164003926 | 36 | 0.355436992 |
| Bone | original_gldm_HighGrayLevelEmphasis | 0.143730756 | -32.2376894 | -1.46203815 | 36 | 0.256860277 |
| Bone | original_gldm_LargeDependenceEmphasis | 0.452292333 | -0.011111413 | -0.751598854 | 36 | 0.525789837 |
| Bone | original_gldm_LargeDependenceHighGrayLevelEmphasis | 0.005604761 | -0.455001613 | -2.770050454 | 36 | 0.015795236 |
| Bone | original_gldm_LargeDependenceLowGrayLevelEmphasis | 0.154668554 | -0.031536981 | -1.423233236 | 36 | 0.256860277 |
| Bone | original_gldm_LowGrayLevelEmphasis | 0.741468814 | -0.026918979 | -0.329908879 | 36 | 0.825606063 |
| Bone | original_gldm_SmallDependenceEmphasis | 0.284185673 | -4.75163707 | -1.070963875 | 36 | 0.367073161 |
| Bone | original_gldm_SmallDependenceHighGrayLevelEmphasis | 0.268459036 | -3476.074833 | -1.106618199 | 36 | 0.367073161 |
| Bone | original_gldm_SmallDependenceLowGrayLevelEmphasis | 0.003623977 | 3.427765491 | 2.909163155 | 36 | 0.01203678 |
| Bone | original_glrlm_GrayLevelNonUniformity | 0.44698483 | 0.015439284 | 0.760451365 | 36 | 0.525789837 |
| Bone | original_glrlm_GrayLevelNonUniformityNormalized | 0.090410116 | 0.059690471 | 1.693238304 | 36 | 0.178896613 |
| Bone | original_glrlm_GrayLevelVariance | 0.254194174 | -243.2974603 | -1.140221161 | 36 | 0.35972909 |
| Bone | original_glrlm_HighGrayLevelRunEmphasis | 0.154042417 | -38.27151837 | -1.425397199 | 36 | 0.256860277 |
| Bone | original_glrlm_LongRunEmphasis | 3.36E-07 | 0.04150898 | 5.101890271 | 36 | 2.61E-06 |
| Bone | original_glrlm_LongRunHighGrayLevelEmphasis | 0.313956755 | -1.033816752 | -1.00695426 | 36 | 0.389306376 |
| Bone | original_glrlm_LongRunLowGrayLevelEmphasis | 0.989881376 | -0.000181126 | -0.012682155 | 36 | 0.989881376 |
| Bone | original_glrlm_LowGrayLevelRunEmphasis | 0.831267761 | -0.017893209 | -0.213075902 | 36 | 0.880639447 |
| Bone | original_glrlm_RunEntropy | 0.019946553 | -0.092485212 | -2.327351719 | 36 | 0.045244621 |
| Bone | original_glrlm_RunLengthNonUniformity | 0.001594351 | -1.581261071 | -3.156938252 | 36 | 0.005702871 |
| Bone | original_glrlm_RunLengthNonUniformityNormalized | 2.61E-06 | -0.778030027 | -4.699244513 | 36 | 1.52E-05 |
| Bone | original_glrlm_RunPercentage | 3.63E-09 | -0.564109453 | -5.900189751 | 36 | 5.54E-08 |
| Bone | original_glrlm_RunVariance | 0.277813622 | 0.007804266 | 1.085243953 | 36 | 0.367073161 |
| Bone | original_glrlm_ShortRunEmphasis | 1.47E-05 | -0.29830677 | -4.333105921 | 36 | 7.60E-05 |
| Bone | original_glrlm_ShortRunHighGrayLevelEmphasis | 0.147054108 | -115.5060629 | -1.450015819 | 36 | 0.256860277 |
| Bone | original_glrlm_ShortRunLowGrayLevelEmphasis | 0.745708702 | -0.047846649 | -0.324302929 | 36 | 0.825606063 |
| Bone | original_glszm_GrayLevelNonUniformity | 0.18440089 | -0.531497649 | -1.327325931 | 36 | 0.295677288 |
| Bone | original_glszm_GrayLevelNonUniformityNormalized | 0.109576963 | 0.072298032 | 1.600097468 | 36 | 0.207972602 |
| Bone | original_glszm_GrayLevelVariance | 0.255291612 | -108.3351295 | -1.137590268 | 36 | 0.35972909 |
| Bone | original_glszm_HighGrayLevelZoneEmphasis | 0.265829098 | -37.34491233 | -1.112719081 | 36 | 0.367073161 |
| Bone | original_glszm_LargeAreaEmphasis | 0.16205881 | 0.001606844 | 1.398180704 | 36 | 0.264411742 |
| Bone | original_glszm_LargeAreaHighGrayLevelEmphasis | 0.771493706 | 0.001518644 | 0.29042162 | 36 | 0.844104878 |
| Bone | original_glszm_LargeAreaLowGrayLevelEmphasis | 0.41680092 | 0.000480877 | 0.811983813 | 36 | 0.496954943 |
| Bone | original_glszm_LowGrayLevelZoneEmphasis | 0.123383007 | 0.098449049 | 1.540727967 | 36 | 0.229492393 |
| Bone | original_glszm_SizeZoneNonUniformity | 0.844167095 | 1.196034468 | 0.196566108 | 36 | 0.880639447 |
| Bone | original_glszm_SizeZoneNonUniformityNormalized | 0.000344017 | 0.651769028 | 3.579678908 | 36 | 0.001523502 |
| Bone | original_glszm_SmallAreaEmphasis | 2.85E-08 | 33.34479913 | 5.5502876 | 36 | 2.65E-07 |
| Bone | original_glszm_SmallAreaHighGrayLevelEmphasis | 0.549778669 | 438.0884461 | 0.598091818 | 36 | 0.631227361 |
| Bone | original_glszm_SmallAreaLowGrayLevelEmphasis | 0.000797372 | 5.697133438 | 3.35370554 | 36 | 0.003011079 |
| Bone | original_glszm_ZoneEntropy | 0.000290337 | -0.367120033 | -3.623772272 | 36 | 0.001350068 |
| Bone | original_glszm_ZonePercentage | 0.046839529 | -15.65770553 | -1.987748348 | 36 | 0.099001731 |
| Bone | original_glszm_ZoneVariance | 0.15234106 | 0.001502266 | 1.43131115 | 36 | 0.256860277 |
| Bone | original_ngtdm_Busyness | 0.036106795 | 0.015506602 | 2.095722725 | 36 | 0.07809144 |
| Bone | original_ngtdm_Coarseness | 0.862997708 | -0.184942823 | -0.172559475 | 36 | 0.88196469 |
| Bone | original_ngtdm_Complexity | 0.291115715 | -1550.273883 | -1.055677206 | 36 | 0.370873445 |
| Bone | original_ngtdm_Contrast | 0.410699633 | -3.758734068 | -0.82266305 | 36 | 0.496039817 |
| Bone | original_ngtdm_Strength | 0.28250776 | -1235.091808 | -1.074702911 | 36 | 0.367073161 |

**Supplementary 04**

| | Lung Only | Mediastinal Only | Bone Only | Lung and Mediastinal | Lung and Bone | Mediastinal and Bone |
|---|---|---|---|---|---|---|
| **Count** | 25 | 3 | 17 | 5 | 20 | 1 |
| **Features** | original_glszm_LowGrayLevelZoneEmphasis, original_glcm_SumSquares, original_glrlm_GrayLevelNonUniformityNormalized, original_glszm_SizeZoneNonUniformity, original_ngtdm_Contrast, original_glcm_ClusterProminence, original_glcm_JointEnergy, original_glrlm_GrayLevelVariance, original_gldm_LargeDependenceEmphasis, original_firstorder_Variance, original_glcm_Id, original_firstorder_Range, original_gldm_DependenceVariance, original_glcm_MaximumProbability, original_glrlm_ShortRunLowGrayLevelEmphasis, original_glszm_HighGrayLevelZoneEmphasis, original_gldm_GrayLevelVariance, | original_gldm_LowGrayLevelEmphasis, original_glcm_Autocorrelation, original_firstorder_Kurtosis | original_gldm_DependenceEntropy, original_gldm_GrayLevelNonUniformity, original_ngtdm_Busyness, original_glcm_Correlation, original_glrlm_RunLengthNonUniformity, original_glszm_SmallAreaLowGrayLevelEmphasis, original_glcm_DifferenceEntropy, original_glcm_SumEntropy, original_glcm_DifferenceAverage, original_glcm_Idmn, original_gldm_SmallDependenceLowGrayLevelEmphasis, original_firstorder_Skewness, original_glszm_ZoneEntropy, original_glcm_JointEntropy, original_glcm_Imc2, original_firstorder_Entropy, original_glcm_MCC | original_glcm_DifferenceVariance, original_glrlm_LongRunHighGrayLevelEmphasis, original_glcm_Contrast, original_gldm_HighGrayLevelEmphasis, original_glrlm_HighGrayLevelRunEmphasis | original_firstorder_Mean, original_firstorder_RootMeanSquared, original_gldm_DependenceNonUniformity, original_firstorder_10Percentile, original_glcm_SumAverage, original_firstorder_InterquartileRange, original_firstorder_90Percentile, original_glrlm_RunEntropy, original_gldm_DependenceNonUniformityNormalized, original_glszm_SizeZoneNonUniformityNormalized, original_firstorder_Minimum, original_firstorder_Median, original_glcm_JointAverage, original_firstorder_Maximum, original_glszm_SmallAreaE | original_gldm_LargeDependenceHighGrayLevelEmphasis |